\documentclass[
aps,prb,twocolumn,superscriptaddress,
floatfix,
showpacs,amsmath,amssymb
]{revtex4-1}

\usepackage{graphicx,color,bbm}
\usepackage{epsfig}
\newcommand{\bra}[1]{\langle{#1} |}
\newcommand{\ket}[1]{|{#1}\rangle  }

\begin{document}

\title{Design of a Lambda system for population transfer in 
superconducting nanocircuits}

\author{G. Falci}
\email[gfalci@dmfci.unict.it]{}
\affiliation{Dipartimento di Fisica e Astronomia,
Universit\`a di Catania, Via Santa Sofia 64, 95123 Catania, Italy}
\affiliation{CNR-IMM  UOS Universit\`a (MATIS), 
Consiglio Nazionale delle Ricerche, Via Santa Sofia 64, 95123 Catania, Italy}
\affiliation{Centro Siciliano di Fisica Nucleare e Struttura della Materia,
 Via Santa Sofia 64, 95123 Catania, Italy}
\author{A. La Cognata}
\affiliation{Centro Siciliano di Fisica Nucleare e Struttura della Materia,
 Via Santa Sofia 64, 95123 Catania, Italy}
\author{M. Berritta}
\altaffiliation[Also at ]{Department of Physics, Lancaster University, 
Lancaster LA1 4YB, United Kingdom.}
\affiliation{Dipartimento di Fisica e Astronomia,
Universit\`a di Catania, Via Santa Sofia 64, 95123 Catania, Italy}
\author{A. D'Arrigo}
\affiliation{CNR-IMM  UOS Universit\`a (MATIS), 
Consiglio Nazionale delle Ricerche, Via Santa Sofia 64, 95123 Catania, Italy}
\author{E. Paladino}
\affiliation{Dipartimento di Fisica e Astronomia,
Universit\`a di Catania, Via Santa Sofia 64, 95123 Catania, Italy}
\affiliation{CNR-IMM  UOS Universit\`a (MATIS), 
Consiglio Nazionale delle Ricerche, Via Santa Sofia 64, 95123 Catania, Italy}
\affiliation{Centro Siciliano di Fisica Nucleare e Struttura della Materia,
 Via Santa Sofia 64, 95123 Catania, Italy}
\author{B. Spagnolo}
\affiliation{Dipartimento di Fisica e Chimica, 
Universit\`{a} di Palermo, Group of Interdisciplinary Physics and CNISM,
Unit\`{a} di Palermo, Viale delle Scienze, Ed.18, I-90128 Palermo, Italy}

\date{\today}

\begin{abstract}
The implementation of a Lambda scheme in superconducting artificial atoms
could allow detection of stimulated Raman adiabatic passage (STIRAP) 
and other quantum manipulations in the microwave regime. However 
symmetries which on one hand protect the system against 
decoherence, yield selection rules which may cancel coupling to the 
pump external drive. The tradeoff 
between efficient coupling and decoherence due to broad-band colored 
Noise (BBCN), which is often the main source 
of decoherence is addressed, in the class of nanodevices based on the 
Cooper pair box (CPB) design. We study 
transfer efficiency by STIRAP, showing that substantial efficiency 
is achieved for off-symmetric bias  only in the charge-phase regime.  
We find a number of results uniquely due to non-Markovianity of BBCN,
namely: (a) the efficiency for STIRAP depends essentially on 
noise channels in the trapped subspace; (b) 
low-frequency fluctuations can be analyzed and represented
as fictitious correlated fluctuations of the detunings of the 
external drives;  
(c) a simple figure of merit for design and operating prescriptions 
allowing the observation of STIRAP is proposed. The emerging physical 
picture also applies to other classes of coherent nanodevices subject to BBCN. 
\end{abstract}



\maketitle

\section{Introduction}
\begin{figure}[pt]
\centering
\resizebox{0.45\textwidth}{!}{\includegraphics{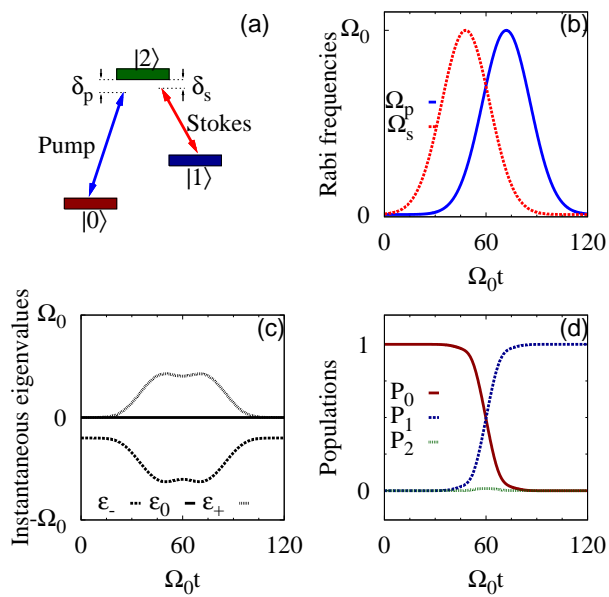}}
\caption{(color online) (a) Three-level system driven with AC fields in  
$\Lambda$ configuration. (b) The 
counterintuitive sequence: the Stokes field 
is switched on {\em before} the pump field (here  
$\Omega_0 T = 20, \tau= 0.6 \,T$).  
(c) Instantaneous eigenvalues $\{\epsilon_0(t),\epsilon_{\pm}(t)\}$, 
for $\delta =0$, $\delta_p= -0.2\,\Omega_0$ and 
$\kappa = 1$.
(d) Population histories 
$\rho_{ii}(t) = |\langle i |\psi(t) \rangle|^2$
for ideal STIRAP ($\delta = 0$): 
the system prepared in $|0 \rangle$ follows 
the Hamiltonian along the $\epsilon_0$ adiabatic path 
yielding complete population transfer to $\ket{1}$.  
\label{fig:stirap-ideal}}
\end{figure}

The rapid technological progress in quantum-state engineering in 
superconducting nanodevices demands for the implementation of 
new advanced techniques of quantum control. 
STIRAP~\cite{kr:198-bergmann-rmp,kr:201-vitanov-annurev} is a powerful 
method in quantum optics,
which is still largely unexplored in the solid-state realm. 
Using AC driving fields in $\Lambda$ 
configuration (see Fig.~\ref{fig:stirap-ideal}.a)
a quantum $M>2$-state system is trapped into a subspace spanned by the 
two longest lived states. Control in 
this trapping subspace can be achieved by adiabatic time evolution induced by properly crafted pulses, allowing for instance to prepare a given 
target state~\cite{kr:196-arimondo-progopt-cpt,kb:197-scully-zubairy}. 
Adiabatic passage used in STIRAP guarantees 
highly efficient and selective population 
transfer in atomic and molecular systems~\cite{kr:198-bergmann-rmp,kr:201-vitanov-annurev}.

In the last few years it has been proposed that 
multilevel quantum coherent 
effects~\cite{kb:197-scully-zubairy}
could be observed in superconducting nanodevices, 
for instance electromagnetically induced transparency 
(EIT)~\cite{ka:204-murali-orlando-prl-eit} or
selective population transfer by STIRAP~\cite{ka:204-siewert-brandes-advsolst-stirap,ka:205-nori-prl-adiabaticpassage,ka:206-siebrafalci-optcomm-stirap,
ka:209-siebrafalci-prb}.
This would be important both from a fundamental point of view, since coherent 
dynamics in multilevel atoms clearly displays beautiful interference 
phenomena~\cite{kb:197-scully-zubairy}, and for applications. These 
include the implementation of microwave 
quantum switches~\cite{ka:212-li-hakonen-nature-qswitch}, 
the manipulation of solid-state qubit 
circuits~\cite{ka:208-wei-nori-prl-stirapqcomp}
and the fascinating perspectives of coupling strongly such nanodevices 
to electromagnetic~\cite{ka:205-mariantoni-aXiv-microwfock,kr:211-you-nori-nature-multilevel} or nanomechanical quantized modes~\cite{ka:209-siebrafalci-prb}.
Very recently few experiments have demonstrated features of 
multilevel coherence in such devices, as the Autler-Townes
(AT) splitting~\cite{ka:209-sillanpa-simmonds-prl-autlertownes,ka:209-baur-wallraff-prl-autlertownes}, EIT~\cite{ka:210-abdumalikov-prl-eit}, 
preparation and measurement of three-state 
superpositions~\cite{ka:210-bianchetti-wallraff-control-tomnography-3ls}, 
dynamical AT control~\cite{ka:212-li-hakonen-nature-qswitch} and 
coherent population trapping~\cite{ka:210-kelly-pappas-prl-cpt}. 

In all the above experiments, except that of 
Ref.~\onlinecite{ka:210-kelly-pappas-prl-cpt}, the multilevel system 
was driven in the {\em Ladder} configuration~\cite{kb:197-scully-zubairy}. 
Indeed in order to implement a $\Lambda$ configuration the device 
Hamiltonian should be strongly asymmetric, which may be 
achieved by a proper external biasing~\cite{ka:205-nori-prl-adiabaticpassage,ka:206-siebrafalci-optcomm-stirap,ka:209-siebrafalci-prb,kr:211-you-nori-nature-multilevel},  
otherwise selection rules prevent to drive efficiently 
the pump transition. However the longest decoherence times in quantum bits 
are achieved by biasing the devices at (or near) parity symmetry points. 
Hence difficulties in implementing a $\Lambda$ configuration
in superconducting nanodevices raise a  
fundamental design issue. 
In particular, low-frequency noise which is known to determine the 
performances of systems operated as 
quantum bits~\cite{kr:208-clarke-wilhelm-nature-squbit} and which is 
minimized at symmetry points, is shown in this work to play 
a major role in degradation of efficiency in STIRAP. 
So far the effect of decoherence in multilevel superconducting 
artificial atoms has been addressed using Markovian master equations. 
In this work we address decoherence effects due to a solid-state environment
where a strong non-Markovian noise component is also present. 
From the exquisite sensitivity of coherence to 
operating conditions, and to design parameters of the device 
we determine the prescriptions for the demonstration 
of a $\Lambda$ scheme in realistic superconducting nanocircuits.

We tackle this problem by a quantitative analysis of  
a class of superconducting nanocircuits, namely those based 
on the Cooper pair box~\cite{ka:198-bouchiat-phscripta-CPB} 
(CPB, see Fig.~\ref{fig:quantronium}). 
This is an important case-study encompassing several 
different coherent nanodevices which have already successfully  
implemented quantum bits~\cite{ka:199-nakamura-nature,ka:202-vion-science,ka:204-duty-delsing-prb-chargeqb,ka:204-wallraff-nature-cqed,ka:207-koch-pra-transmon}. The emerging physical picture is even more general 
holding true for nanodevices suffering mainly from the presence 
of low-frequency noise.

The main message of this work is twofold. First we 
find that observation of STIRAP should be 
possible with devices fabricable at present days, provided that
operating conditions {\em and} suitable design 
optimize the conflicting requirements of efficient coupling between 
states with (approximately) the same  parity
and protection from low-frequency noise. Second, 
despite of the complicated multilevel structure and of the many 
parameters involved, we show that the efficiency 
for STIRAP depends essentially on noise channels
involving the trapping subspace, and determine a simple figure of merit 
for design and operating prescriptions of devices allowing the
observation of STIRAP.

The paper is organized as follows. We introduce STIRAP 
in Sec.~\ref{sec:co-po-tra} and describe population 
transfer via adiabatic and nonadiabatic patterns. 
In Sec.~\ref{sec:stirap-quant} we discuss the implementation in a CPB, 
and introduce the model for broad-band colored noise (BBCN), extending 
to a $\Lambda$ system the approach introduced in 
Refs.~\onlinecite{ka:205-falci-prl,ka:falci-varenna}, 
which quantitatively explains qubit decoherence due to 
BBCN in superconducting 
qubits~\cite{ka:202-nakamura-chargenoise,ka:205-ithier-prb,ka:211-bylander-natphys,ka:212-chiarello-njp,ka:212-sank-martinis-prl-fluxnoise}. In Sec.~\ref{sec:results-broadband} we present results on the 
effects of the BBCN, focusing on the charge-phase regime of CPB's.
In Sec.~\ref{sec:optimal-design} we extend the above considerations to other 
regimes of the CPB and determine the figure of merit characterizing 
optimal design and operating conditions.
In Sec.~\ref{sec:comparison} we compare the effects of
dephasing with long memory time with Markovian dephasing, showing that in
the former case driving more strongly the system would improve efficiency.
Therefore STIRAP could in principle discriminate between different 
dynamic characteristics of decoherence sources 
in superconducting nanocircuits. 
Conclusions are drawn in Sec.~\ref{sec:conclusion}.

\section{Coherent population transfer in three-level atoms}
\label{sec:co-po-tra}

\subsection{Dark state and STIRAP}
In quantum optics STIRAP is based on a 
$\Lambda$ configuration (Fig.~\ref{fig:stirap-ideal}a) 
of two hyperfine ground states
$|0\rangle$ and  $|1\rangle$ and an excited state
$|2\rangle$, with energies $E_{0}=0$, $E_{{1}}$ and 
$E_{{2}}$ respectively.
The system is operated by two classical laser 
fields~\cite{kr:198-bergmann-rmp,kr:201-vitanov-annurev,kr:196-arimondo-progopt-cpt,kb:197-scully-zubairy}, 
the Stokes laser $\Omega_{12} = \Omega_s \cos{\omega_s t}$ 
and the pump laser $\Omega_{02}=\Omega_p \cos{\omega_p t}$, each 
being nearly resonant with the corresponding transition.
The effective Hamiltonian is conveniently written in a doubly rotating frame
at the angular frequencies $\omega_k$, where $k=p,s$ refer to 
pump and Stokes. In the rotating-wave approximation (RWA)
the effective Hamiltonian is~\cite{kr:198-bergmann-rmp,kb:197-scully-zubairy,kr:201-vitanov-advmolopt} 
\begin{equation}
\tilde{H}= 
\delta |1\rangle\langle 1| +
\delta_{{p}} |2\rangle\langle 2| 
           + \Big( \frac{\Omega_s}{2} |2\rangle\langle 1|
           + \frac{\Omega_p}{2} |2\rangle\langle 0|+{\rm h.c.}\Big) \ 
\label{eq:lambdaconfig}
\end{equation}
Here 
$\delta_s=E_{2}-E_{1}-\omega_s$ and $\delta_p=E_{2}-E_{0}-\omega_p$
are the single-photon detunings and we introduced the
two-photon detuning 
$\delta=\delta_p-\delta_s 
$. Both the detunings and the Rabi frequencies $\Omega_k$ can be 
functions of time.  
At two-photon resonance, $\delta =0$, 
the Hamiltonian (\ref{eq:lambdaconfig}) 
has a zero energy instantaneous eigenvalue $\epsilon_0 =0$
(Fig.~\ref{fig:stirap-ideal}c)
whose eigenstate is a ``dark state'', 
\begin{equation}
         |D \rangle\ =\ 
         \frac{\Omega_s |0\rangle\ -\ \Omega_p|1\rangle}
         {\sqrt{|\Omega_s|^2 + |\Omega_p|^2}}\,                 
\label{eq:darkclass}
\end{equation}
and two other eigenstates $\ket{\pm}$ with nonzero eigenvalues 
$\epsilon_\pm = {1 \over 2} \,\delta_p \pm {1 \over 2} 
\sqrt{\delta_p^2 + \Omega_s^2 + \Omega_p^2}$
whose form can be found analytically~\cite{kr:198-bergmann-rmp}. 
If the system is in the dark state, the population is trapped in the two 
lowest diabatic states $\{\ket{0},\ket{1}\}$. This is due to 
destructive interference of the two 
fields: despite excitation by the lasers the state $|2\rangle$ 
is never populated and no radiative decay can be detected.

By slowly varying the coupling strengths, $\Omega_s(t)$ and $\Omega_p(t)$, 
the dark state can be rotated adiabatically in the subspace spanned 
by $|1\rangle$ and $|0\rangle$. In particular 
STIRAP yields complete coherent population transfer 
$|0\rangle \to |1\rangle$ as follows~\cite{kr:198-bergmann-rmp}:
the system is prepared in $|0\rangle$, which coincides with the dark state 
for $\Omega_s=\Omega_p=0$; then $\Omega_s$ is slowly switched on; 
after a delay $\tau$ also $\Omega_p$ is slowly switched on; at this stage
$\Omega_s$ is slowly switched off and 
the dark state now coincides with $|1\rangle$; finally the protocol ends
by switching off $\Omega_p$, achieving complete population transfer 
(Fig.~\ref{fig:stirap-ideal}.d). 
Notice that in STIRAP 
population transfer is achieved by a ``counterintuitive'' 
pulse sequence (Fig.~\ref{fig:stirap-ideal}.b), 
which has several 
advantages~\cite{kr:198-bergmann-rmp,kr:201-vitanov-advmolopt}. 
First the excited state $|2\rangle$, which may undergo strong 
spontaneous decay 
deteriorating the transfer efficiency, is never populated during STIRAP. 
Moreover, provided adiabaticity is preserved,  
STIRAP is insensitive to many details 
of the protocol, as the precise timing of the operations, 
a property which makes it attractive for implementing
fault-tolerant quantum 
gates~\cite{ka:208-wei-nori-prl-stirapqcomp,ka:211-timoney-nature-dressed}.

\begin{figure}[pt]
\centering
\resizebox{0.35\textwidth}{!}
{\includegraphics{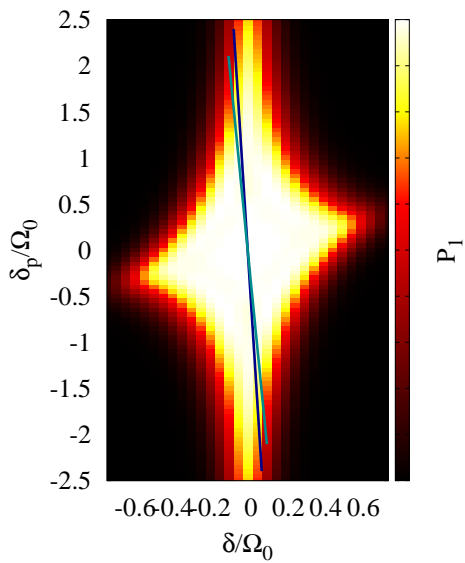}}
\caption{(color online) Sensitivity to detunings of the efficiency of STIRAP. 
Here 
$\Omega_0 T = 20$, $\tau = T/2$ and $\kappa=1$. In the white 
zone the efficiency is larger than 90\%. 
Efficiency is very sensitive to a nonzero 
two-photon detuning $\delta$, and much less sensitive to 
$\delta_p \neq 0$ (notice the different scale of the 
axis). 
The two lines on the plot represent correlated stray detunings 
induced in the CPB by charge noise 
for two different values of $q_g=0.47,0.49$
(see Sec.~\ref{sec:effmodel-lowfreq}, and~\ref{sec:lowfreq-quantronium}). 
\label{fig:stirap-diamond}
}
\end{figure}

\subsection{Sensitivity to parameters}
\label{sec:2}
Adiabaticity is important to achieve high efficiency since non adiabatic 
effects trigger unwanted transitions detrapping the system from the dark 
state. A necessary condition for adiabaticity is~\cite{kr:198-bergmann-rmp}
$|(\dot{\Omega}_p \Omega_s - \Omega_p \dot{\Omega}_s )/ 
(\Omega_p^2 + \Omega^2_s )|  \ll |\epsilon_\pm-\epsilon_0|$
which suggests that large enough 
Rabi peak angular frequencies $\Omega_k$ are needed, 
in order to determine a large AT 
splitting of the instantaneous 
eigenstates.
We let
$\Omega_p(t) = \Omega_0 \,f[(t-\tau)/T]$ and 
$\Omega_s(t) = \kappa \Omega_0 \,f[(t+\tau)/T]$.
A positive delay $\tau$ implements the 
counterintuitive sequence. 
For Gaussian pulses, $f(x)=\mathrm{e}^{-x^2}$, the 
choice $\tau > (\sqrt{2}-1)\, T$ and $\Omega_0 T \gg 10$ yields 
efficient population transfer~\cite{kr:201-vitanov-advmolopt}. 
As we discuss later, in superconducting nanocircuits 
the pump peak Rabi angular frequency $\Omega_0$ cannot be very large, and 
$T$ is limited by decoherence. 
We found a good tradeoff for 
$\Omega_0 T= 15$ and a delay $\tau=0.5\,T$,
which turns out to be a satisfactory choice when fluctuations 
of parameters are considered, and which we use unless otherwise specified.

\begin{figure}[pt]
\centering
\resizebox{0.23\textwidth}{!}{\includegraphics{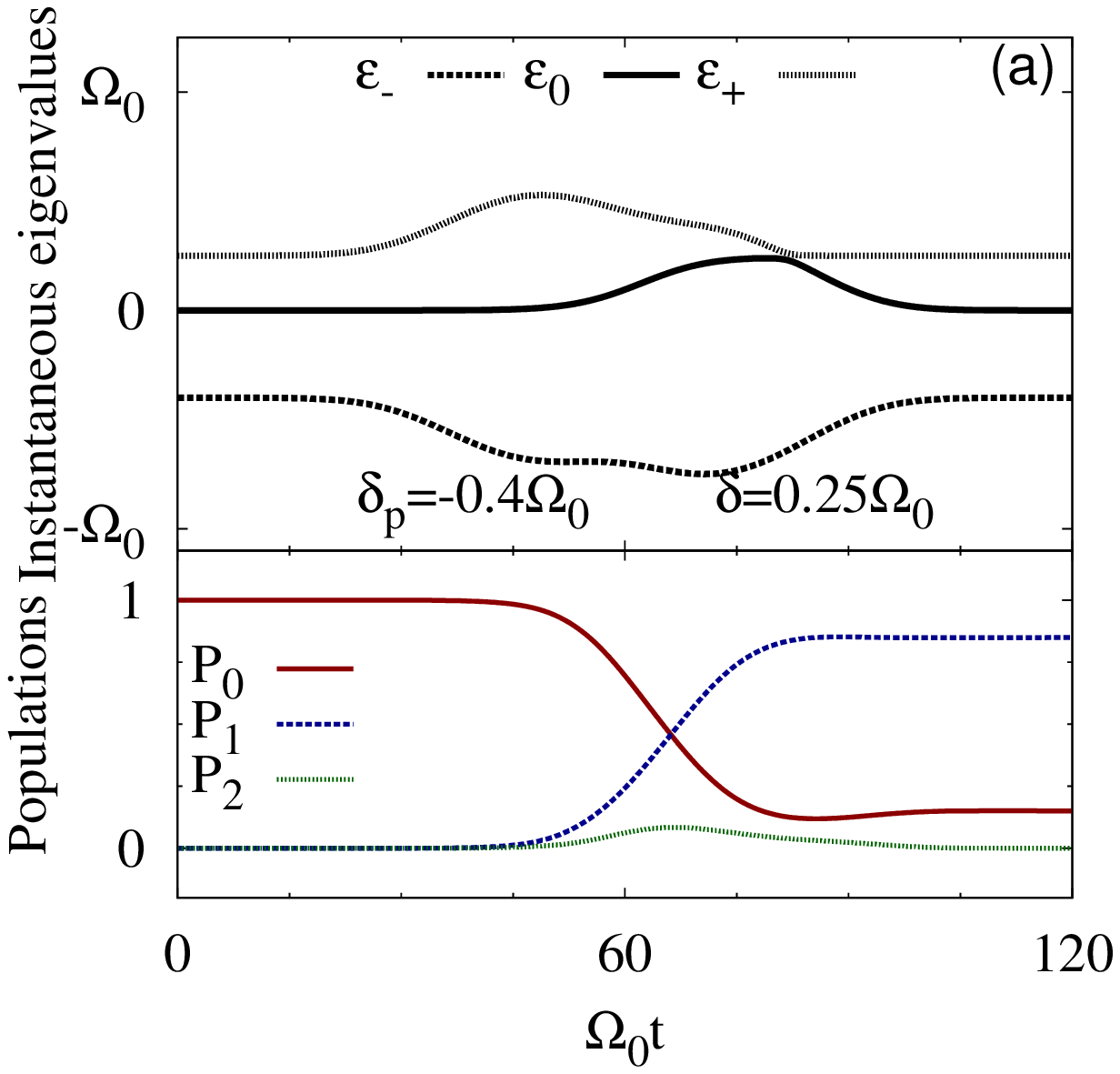}}
\resizebox{0.23\textwidth}{!}{\includegraphics{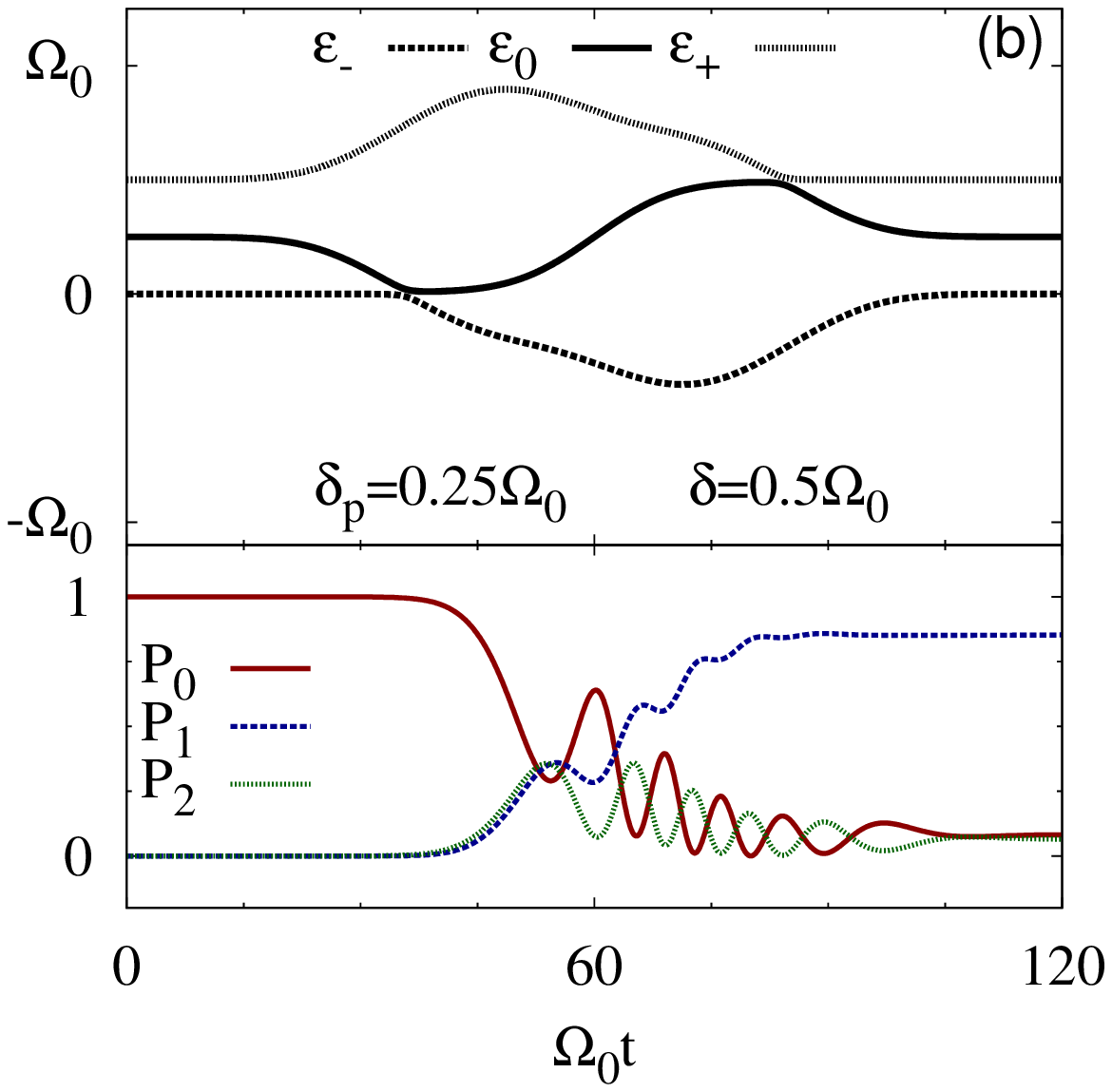}}
\hfill
\caption{(color online) Non-ideal STIRAP ($\Omega_0 T = 20, \tau= 0.6\,T$) 
with 
$\delta \neq 0$, shows different classes of patterns of 
instantaneous eigenstates.
(a) Top: instantaneous eigenstates 
for $\delta = 0.25\,\Omega_0$ and 
$\delta_p = - 0.4 \,\Omega_0$. 
These LZ patterns with a single 
avoided crossing during the pump induced EIT phase
result from the effect of low-frequency charge noise 
in CPB's (or flux noise in flux 
qubits).
Bottom: population histories 
of the diabatic states. 
(b) Top: generic LZ pattern ($\delta=0.5\,\Omega_0$,  
$\delta_p=0.5\,\Omega_0$); bottom: population histories. 
\label{fig:stirap-zener}}
\end{figure}

Non-zero detunings $\delta_s$ and $\delta_p$ modify the 
whole adiabatic picture of STIRAP and may strongly 
affect the transfer efficiency. 
The crucial parameter is the two-photon detuning 
since for $\delta \neq 0$ the dark state (\ref{eq:darkclass}) 
is not anymore an instantaneous eigenstate and  
there is no adiabatic connection from the initial to the target state. 
As a consequence the efficiency is very sensitive to fluctuations
of the two-photon detuning $\delta$, whereas large single-photon 
detunings $\delta_p$ are tolerable (see Fig.~\ref{fig:stirap-diamond} and 
the discussion in Ref.~\onlinecite{kr:201-vitanov-advmolopt}).
For $\delta \neq 0$ the simple picture of adiabatic passage is not valid
anymore and qualitatively new phenomena occur enriching the physical scenario.
In particular non-ideal STIRAP may still take place via non-adiabatic 
transitions between adiabatic states. 
For small values of $\delta$, narrow avoided 
crossings between the instantaneous eigenvalues occur 
and the population is transferred by 
Landau-Zener (LZ) tunneling~\cite{kr:201-vitanov-advmolopt}
(see Fig.~\ref{fig:stirap-zener}). For increasing $\delta$ the transfer 
efficiency is reduced and in general the excited state $|2\rangle$ 
is populated during the protocol.

It is worth stressing the importance of correlations between detunings.
Indeed it is well known in atomic 
physics~\cite{ka:202-yatsenko-pra-stirapphasenoisecorr} that   
if $\delta_s$ and $\delta_p$ are correlated as to nearly preserve  
two-photon resonance, still a large transfer efficiency is obtained.
In superconducting nanodevices correlations
of other nature may arise between effective fluctuations of 
$\delta$ and $\delta_p$, induced by solid-state noise. 
These correlations are represented by the lines in 
Fig.~\ref{fig:stirap-diamond}, 
and determine the typical pattern for population transfer via 
LZ processes~\cite{ka:212-falci-phscripta-opendriven}
of Fig.~\ref{fig:stirap-zener}a. Notice that in this case relatively large 
single-photon detunings $\delta_p \sim 25 \,\delta$ still allow 
coherent population transfer. This has important consequences in 
coherent nanodevices where fluctuations may produce large detunings 
$\delta_p$.  

\begin{figure}[pt]
\centering
\resizebox{0.27\textwidth}{!}{\includegraphics{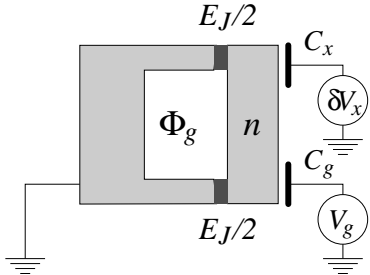}}
\caption{In the CPB design the state of the superconducting island 
is a superpositions of states with well defined 
number $n$ of extra Cooper pairs.
The device is biased by the gate voltage $V_g$ determining 
the operating point of $q_g=V_g/2 e C_g$; 
control is operated by an ac component of $V_g$. 
Charge fluctuations are equivalent to voltage fluctuations $\delta V_x$.
The effective Josephson energy can be tuned via the flux $\Phi_g$ 
of the magnetic field threading 
the loop, $E_J=E_J(\Phi_g)$. 
\label{fig:quantronium}}
\end{figure}

\section{STIRAP in the Cooper pair Box}
\label{sec:stirap-quant}

\subsection{Implementation of the $\Lambda$ system}
\begin{figure}[pt]
\centering
\resizebox{!}{0.25\textheight}{\includegraphics{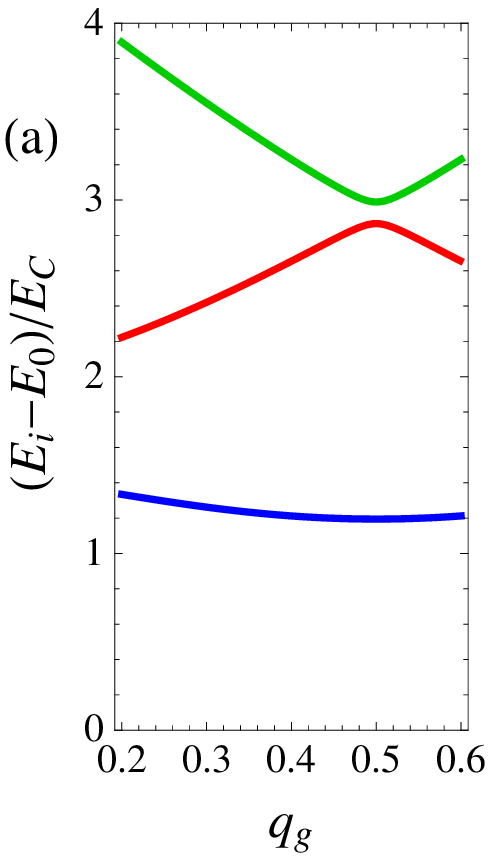}}
\resizebox{!}{0.26\textheight}{\includegraphics{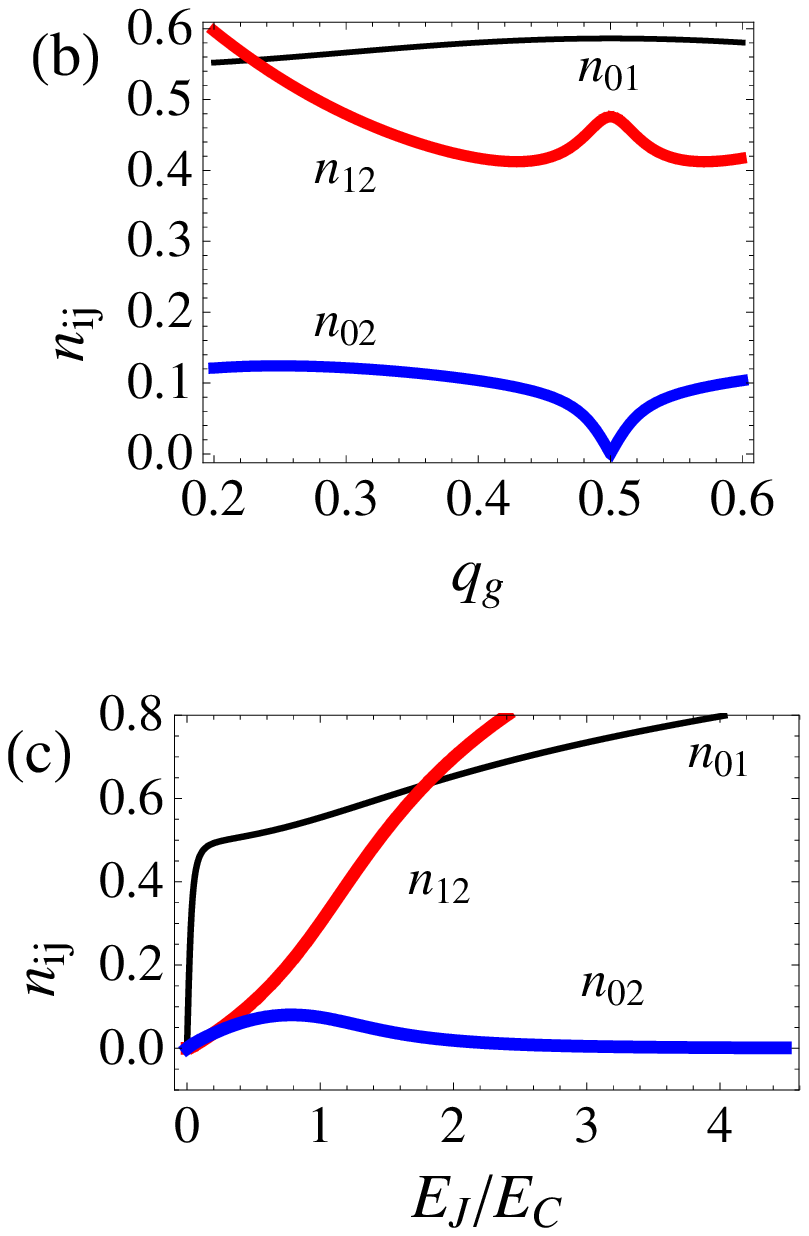}}
\caption{(color online)
(a) Energy spectrum $E_i$ of a charge-phase 
CPB for $J=1.32$ 
(corresponds to the 
Quantronium~\cite{ka:202-vion-science}), 
relative to the ground state $E_0=0$,  
vs. the bias $q_g$.   
(b)~Matrix elements of $\hat{n}$ involved in the 
$\Lambda$ scheme vs. $q_g$ for 
$J = 1.32$; the element $n_{02}$ vanishes at the symmetry
point $q_g=1/2$; (c)~Matrix elements vs  $J=E_J/E_C$ for 
$q_g=0.48$; notice that $n_{02}$ is 
much smaller than other elements (it vanishes at $q_g=1/2$) 
and it has nonmonotonous behavior for increasing $J$.
\label{fig:spectrum}
}
\end{figure}

The CPB~\cite{ka:198-bouchiat-phscripta-CPB}
is a superconducting loop interrupted by two adjacent small 
Josephson junctions (energy $E_J/2$) defining a superconducting island 
(Fig.~\ref{fig:quantronium}). The total 
capacitance $C$ gives the charging energy $E_C=(2e)^2/ 2 C$. 
The electrostatic energy is modulated by a gate voltage $V_g$, 
connected to the island via a capacitance $C_g \ll C$.
The Hamiltonian reads
\begin{equation}
\label{eq:quantronium_hamilt}
H_0(q_g) = \sum_n E_{{C}} (n-q_g)^2|n\rangle\!\langle n|
- \frac{E_{J}}{2} (|n\rangle\!\langle n+1| +\mbox{h.c.})
\end{equation}
where $\{|n\rangle,\, n\in]-\infty,\infty[\}$ 
are eigenstates of the number operator $\hat{n}$ of extra Cooper pairs 
in the island. We have defined the reduced gate charge 
$q_g= C_g V_g/(2e)$ polarizing the island. 
The spectrum can be modified by choosing a specific bias $q_g$ 
(Fig.~\ref{fig:spectrum}). 

The parametric dependence of $H_0$ on $q_g$  
defines a port allowing for external control of the system:
by adding an ac microwave component $q_g \to q_g+q_c(t)$, shaped in suitable 
pulses, arbitrary rotations of the quantum state have been 
demonstrated~\cite{ka:204-collin-saclay-prb-nmrmanipulation}.

In the basis of the eigenvectors $\{|\phi_i(q_g)\rangle,\, i=0,1,2\}$
of $H_0(q_g)$ the driven Hamiltonian reads
\begin{equation}
\label{eq:quantronium_hamilt-ac}
H(t) = \sum_i E_i 
|\phi_i\rangle \langle \phi_i|
+ A(t) \, \sum_{ij} n_{ij} \,|\phi_i\rangle \langle \phi_j|
\end{equation}
where $n_{ij}=\langle \phi_i | \hat{n} |\phi_j \rangle$
and the control field is $A(t) = - 2 E_C \,q_c(t)$.   
For STIRAP we let 
$A(t)={\cal A}_s(t)\cos{\omega_s t}+{\cal A}_p(t)\cos{\omega_p t}$. 
We then transform the Hamiltonian 
to the doubly rotating frame, and retain only slowly varying terms, which
yields the RWA (see App.~\ref{app:cpb-effhamil}). 
By projecting onto the three
lowest levels, $i,j=0,1,2$, we finally obtain 
an effective Hamiltonian $\tilde{H}$ implementing the $\Lambda$ 
configuration of Eq.(\ref{eq:lambdaconfig}), with the 
definitions
\begin{equation}
\label{eq:rabi-frequencies} 
\Omega_p = n_{02}\,{\cal A}_p \quad ; \quad \Omega_s = n_{12}\,{\cal A}_s.
\end{equation}
Therefore, $\hat{n}$ enters the peak 
Rabi angular frequencies,  as the electric 
dipole does in atoms. The CPB is a tunable atom since 
the parametric dependence 
on $q_g$ (see Eq.(\ref{eq:quantronium_hamilt})) affects 
``diabatic states'', eigenenergies, and matrix elements of 
$\hat{n}$ (see Fig.~\ref{fig:spectrum}). 
Therefore, detunings  and 
peak Rabi frequencies in $\tilde{H}$ 
Eq.(\ref{eq:lambdaconfig}) depend on $q_g$.

Several superconducting qubits are based on the CPB. 
From the point of view of the model they differ for the values of the 
parameter $J=E_J/E_C$. Computational states, 
which are eigenstates of $H_0$, are superpositions 
of a number of ``charge states'' $|n\rangle$, increasing with 
$J$, and therefore these devices have very different energy spectra. 
Coherent dynamics has been observed in the charge 
regime~\cite{ka:199-nakamura-nature,ka:204-duty-delsing-prb-chargeqb} 
$J \ll 1$, in the charge-phase 
regime~\cite{ka:202-vion-science,ka:204-wallraff-nature-cqed}
$J \sim 1$, and in the phase regime~\cite{ka:207-koch-pra-transmon}
$J \gg 1$ (from several tens up to several hundreds).
Physically these devices greatly differ 
both in the design (size, on-chip readout scheme)
and in characteristics (ease of coupling to control fields,
resilience to noise), these features being crucial for functionality. 
Therefore the CPB allows for a thorough discussion 
of requirements to observe STIRAP in a wide class 
of nanodevices.

Notice finally that, besides transitions to higher energy levels, 
the external field 
coupling to artificial atoms may in principle trigger the $0 \to 1$ 
transition and for $q_g \neq 1/2$ also provides a time-dependent 
diagonal contribution to the effective Hamiltonian. 
We have shown in previous works that this 
has no effect for CPB's in the charge-phase regime, due to the large 
anharmonicity of the spectrum, even in the presence of markovian noise~\cite{ka:206-siebrafalci-optcomm-stirap,ka:209-siebrafalci-prb,ka:208-manganosiefalci-epj-stirap}. Therefore we will safely study the effect of noise in the 
lowest three-level subspace. 
Leakage from this subspace is expected for $J \gg 1$, 
but in this regime STIRAP does not occur
even in the three-level approximation, as we show later.

\subsection{Symmetries, decoherence, selection rules}
Tunability with $q_g$ has been exploited to find optimal 
points where qubit operations are well protected from low-frequency 
noise~\cite{ka:202-vion-science,ka:205-falci-prl}.
For instance for $q_g=1/2$ the 
Hamiltonian (\ref{eq:quantronium_hamilt}) is symmetric for charge-parity 
transformations (App.~\ref{app:cpb-parity}).  
Due to this fact CPB-based qubits biased at symmetry are well 
protected against external noise. This has allowed to obtain 
experimental dephasing times of several hundreds of nanoseconds in 
charge-phase devices~\cite{ka:202-vion-science,ka:205-ithier-prb} 
and ranging from $T^*_2 > 2\,\mu\mathrm{s}$ 
$J \sim 50$ in the phase regime~\cite{ka:207-koch-pra-transmon} 
up to $T_2^* \sim 0.1\,\mathrm{ms}$ 
recently reported~\cite{ka:212-rigetti-steffen-prb-trasmonshapphire}.
At the same time symmetry enforces a selection rule 
preventing transitions between states with the same 
charge-parity. In particular $n_{02}$ vanishes at 
$q_g=1/2$ (see Fig.~\ref{fig:spectrum}b) 
therefore it is not possible to 
implement the $\Lambda$ configuration of 
Eq.(\ref{eq:lambdaconfig}), since $\Omega_p=0$. 
In Refs.~\onlinecite{ka:206-siebrafalci-optcomm-stirap,ka:209-siebrafalci-prb} 
it has been proposed to 
overcome this problem by working slightly off-symmetry
(see Fig.~\ref{fig:spectrum}b,c), and it has been shown 
that the full multilevel structure of a CPB with $E_J = E_C$ allows for 
coherent population transfer 
for $q_g \approx 0.47$, in the presence of Markovian noise.  

We stress that protection from noise and selection rules
are related since they both stem from charge-parity symmetry. 
Notice that increasing $J$ enforces the (approximate)
selection rule in a larger and larger neighborhood of 
the symmetry point $q_g = 1/2$, since it makes less effective 
symmetry breaking terms (asymmetric charging energy). For instance 
Fig.~\ref{fig:spectrum}c shows that 
$n_{02}$ at off-symmetry ($q_g = 0.48$) 
eventually decreases for increasing $J$, making impossible the 
implementation of the $\Lambda$ scheme.

\subsection{Model for charge noise}
In principle each port of the device also allows injection of noise and 
provides channels for decoherence.  
The control port associated to $q_g$ couples to charge noise and in  
this paper we focus on it, since it is the main 
source of low-frequency noise in the CPB for the regimes in which STIRAP 
could be observed.  The structure of the coupling to noise 
can be obtained by allowing for fluctuations of the gate charge
in the Hamiltonian Eq.(\ref{eq:quantronium_hamilt}). 
Their physical origin, besides voltage fluctuations of the circuit have 
been recognized as the effect of switching 
impurities~\cite{ka:202-nakamura-chargenoise,ka:202-paladino-prl-backcharges,other-backcharges}
located in the oxides or in the substrate close to the device. 
We let $q_g \to q_g + x$, where $x$ describes 
stray electrical polarization of the island,  
and write the resulting Hamiltonian as $H=H_0(q_g)+ H_{RW}(t) + \delta H$. 
Here $H_{RW}(t)$ is the control Hamiltonian in the RWA, 
Eq.(\ref{eq:adiab-pass-charging-rwa}), 
whereas 
$\delta H = - 2 E_C\,x \,\hat{n}$ describes fluctuations. 
The structure of coupling to a quantum environment is obtained 
on a phenomenological level by 
``quantizing'' noise. This is obtained by letting 
$\delta H = \hat{X} \,\hat{n} + H_{R}$, where 
$\hat{X}$ is an environment operator and 
$H_{R}$ describes the environment alone, and 
suitable counterterms~\cite{kb:193-weiss-buch}.
Markovian noise can then be studied by deriving a weak coupling 
quantum optical master equation (ME).
However, noise in the solid state has large low-frequency components 
invalidating the ME. A multistage 
approach has been proposed~\cite{ka:205-falci-prl} where 
high and low-frequency noise are separated, the 
latter being approximated by a classical random field. 
Formally $\hat{X} \to  \hat{X}_f - 2 E_C \,x(t)$ 
where $\hat{X}_f$ describes fast environmental quantum degrees of freedom and 
$x(t)$ is a slow classical stochastic process.  
If we let $q_x(t) = q_g + x(t)$ the Hamiltonian is written as 
\begin{equation}
\label{eq:noise-quantronium}
H=H_0[q_g+x(t)]+ H_{RW} (t) + \hat{X} \,\hat{n} + H_{env}
\end{equation}
In many cases low-frequency noise has $1/f$ spectrum 
and the leading contribution of the slow dynamics of $x(t)$ is captured 
by a static-path approximation (SPA) 
i.e. approximating the stochastic process by a suitably distributed random 
variable~\cite{ka:205-falci-prl,ka:205-ithier-prb} $x$. 
In this simpler scenario one should first 
calculate the reduced density matrix 
$\hat{\rho}(t|x)$ for a given stray bias $x$
obtained by tracing out high-frequency (quantum) noise, and then average 
over the distribution $p(x)$. In particular population histories are given by
$P_{i}(t) = \int dx \,p(x)\,\rho_{ii}(t|x)$. Notice that for each 
realization $x$ of the random variable 
the system is prepared and measured in the eigenbasis of 
$H_0(q_g+x)$, which is then conveniently 
used to represent $\rho_{ii}$.

In the case of many weakly coupled 
noise sources $p(x)$ is a Gaussian with standard deviation $\sigma_x$.
The low-frequency noise affects the dynamics via 
fluctuations of energy it induces. 
This point of view provides a simple argument explaining why the symmetry 
point $q_g=1/2$ is well protected. Indeed, since 
at this working point the energy splitting $E_1$ depends only 
quadratically on the fluctuations $x$, energy fluctuations are 
suppressed. Therefore superpositions of the two lowest energy levels 
keep coherence for a longer dephasing time, with 
only a power law suppression of the 
signal~\cite{ka:205-falci-prl,ka:205-ithier-prb}. 
This case is referred as ``quadratic noise'' regime, 
to make a distinction with ``linear noise'' conditions, occurring  
for off-symmetry bias, where energy fluctuations are linear in $x$ 
yielding much stronger decoherence (Gaussian decay law). 

This approach has quantitatively explained the power law decoherence 
observed not only in CPB's~\cite{ka:205-ithier-prb}, 
but also in flux qubits~\cite{ka:211-bylander-natphys} and 
has allowed to find optimal operating point in ultrafast driven 
phase qubits~\cite{ka:212-chiarello-njp}.  
Recently it has been used to discuss properties of multiqubit 
systems~\cite{ka:210-paladino}. The present extension to a $\Lambda$ system 
of the approach of Ref.~\onlinecite{ka:205-falci-prl}
enlightens the role of correlations between detunings, and provides a tool for 
optimal device design.

\subsection{Effective model for low-frequency noise in $\Lambda$ configuration} 
\label{sec:effmodel-lowfreq}
In order to study STIRAP the Hamiltonian (\ref{eq:noise-quantronium}) 
is projected 
onto the subspace spanned by the three lowest energy adiabatic 
eigenvectors 
of $H_0[q_g+x(t)]$. In doing so we assume the adiabaticity of 
the dynamics induced by $x(t)$, which allows to neglect effects 
of the time-dependence of the eigenvectors. Of course in the SPA 
adiabaticity of noise is automatically verified. 
The system plus drive Hamiltonian $H_0[q_g+x(t)]+ H_{RW}(t)$ in the 
rotated frame has the same structure of Eq.(\ref{eq:lambdaconfig}), but 
depends on the realization of the stochastic process.
Fluctuations of the eigenenergies translate into fluctuations of the 
detunings (we let $E_0=0$). In the SPA we have
\begin{equation}
\delta(x) = E_1(q_g+x)- \omega_p + \omega_s
\; ; \; 
\delta_p(x) = E_2(q_g+x) -  \omega_p
\label{eq:det-fluct}
\end{equation}
It is worth stressing that also the effective drive fluctuates, 
via the charge matrix element, for instance 
$\Omega_p=n_{02}[q_g+x(t)]\,{\cal A}_p$. Thus the effect of low-frequency noise
in solid-state devices is conveniently recast in terms of sensitivity 
of the protocol to fictitious imperfections (in both phase and amplitude) 
of the drive. 
This allows to apply to solid state devices 
several results from the quantum optics realm.
For instance, the known critical sensitivity to two-photon detuning, 
translates in the fact that the main figure to be minimized, in order to
achieve efficient population transfer in nanodevices, are  
fluctuations of the lowest energy splitting. This is a quantity which is well 
characterized from the qubit 
dynamics~\cite{ka:205-ithier-prb,ka:211-bylander-natphys,ka:212-chiarello-njp,ka:212-sank-martinis-prl-fluxnoise}.

\section{Effect of broadband colored noise on STIRAP in a charge-phase CPB}
\label{sec:results-broadband}
We now apply the above approach to analyze 
STIRAP in a CPB in the charge-phase regime $E_J \sim E_C$.
An important point is that while dephasing is minimized 
by operating at the symmetry point $q_g=1/2$, the selection rule $n_{02}=0$
apparently prevents to implement STIRAP. 
Therefore it has been proposed to operate slightly 
off-symmetry~\cite{ka:209-siebrafalci-prb,ka:205-nori-prl-adiabaticpassage}, 
where on the other hand decoherence due to low-frequency noise 
increases~\cite{ka:205-ithier-prb}. 
This opens the question of the tradeoff between efficient coupling 
of the driving fields 
and dephasing due to slow excitations in the solid-state.

Since it is convenient to work with the largest possible pump 
Rabi peak frequency $\Omega_0$, we will consider its value as a scale. 
For a given peak value of ${\cal A}_p(t)$ it can be estimated as 
$\Omega_0 = \Omega_R \, n_{02}(q_g)/n_{01}(1/2)$ 
where $\Omega_R$ is the maximal angular frequency for Rabi oscillations 
between the lowest doublet at the symmetry point, which is well characterized
in experiments. We will use frequencies 
corresponding to $\nu_R = 600\,\mathrm{MHz}$, which are 
in principle achievable\cite{kp:vion} 
even if there may be technical problems 
in specific devices. For the Quantronium at 
$q_g=0.48$ this would correspond to 
a maximum $\nu_p = 55\, \mathrm{MHz}$. 

Close to the 
symmetry point coupling of the field with the Stokes transition is larger. 
Therefore, we could easily choose 
$\kappa = \nu_s/\nu_p \approx n_{12}/n_{02} \gg 1$.
However, using larger values of $\nu_s$ does not improve the 
transfer efficiency in 
CPB's~\cite{kr:201-vitanov-advmolopt,ka:211-lacognata-ijqi-stirapcpb},  
therefore we will let $\kappa=1$ hereafter.

\begin{figure}[t]
\centering
\resizebox{0.3\textwidth}{!}{\includegraphics{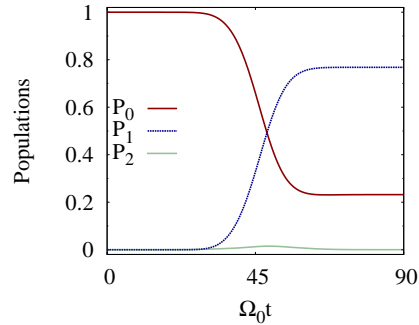}}
\caption{(color online)
Population histories 
in the Quantronium at $q_g=0.475$,  
averaged over fluctuations with  $\sigma_x = 0.004$.  
Charge fluctuations determine anticorrelated stray detunings, 
$\delta_p = - 23.5\, \delta$. Drives are symmetrized, 
$\kappa =1$. The total time for the Quantronium corresponds to 
$\sim \,250\,\mathrm{ns}$. The resulting efficiency is $P_1(t_f) = 0.77$.
\label{fig:avetransf}
}
\end{figure}

\subsection{Effects of low-frequency noise}
\label{sec:lowfreq-quantronium}
Low-frequency fluctuations $x$ of the gate charge 
determine non-exponential dephasing in qubits~\cite{ka:205-falci-prl}. 
They have been well characterized in the Quantronium 
by Ramsey interferometry at different bias points 
$q_g \in [0.4, 0.5]$~\cite{ka:205-ithier-prb}.
Gate charge fluctuations $\sigma_x$ are obtained by the 
measured charging energy fluctuations
$\sigma_E = 2 E_C \sigma_x \sim 0.01\,E_1(1/2)$, a figure 
which is independent on the bias, corresponding to 
$\sigma_x = \sigma_E/ (2 E_C) \approx 6 \cdot 10^{-3}$. These quantities 
are related to the integrated spectral density of the 
environment~\cite{ka:205-falci-prl}, 
and for $1/f^\alpha$ noise they also depend on details of the protocol 
as the total measurement time. 
Even if this dependence is only logarithmic, one can take advantage 
from the fact that measuring the final population in STIRAP requires 
a lower statistics than Ramsey fringes. 
Therefore, for our purposes lower values of $\sigma_x$ are well 
reasonable and hereafter we use $\sigma_x  = 0.004$. This is 
a realistic figure not only for the Quantronium but for the whole class 
of CPB-based devices, since charge noise is ultimately determined by 
material issues which are constantly under investigation.

We consider STIRAP for the optimal conditions of 
nominal single and two-photon resonance, 
$\delta=\delta_p= 0$. 
According to Eq.(\ref{eq:det-fluct})  fluctuations $x$ determine a 
distribution 
of stray detunings. For small $\sigma_x$ we can approximate 
$$
\delta(x) \approx A_1 \,x + {1 \over 2} \, B_1 \, x^2
$$
where $A_1 = (\partial E_1/\partial q_g) = E_C\,a_1(q_g,J)$ 
and $B_1= (\partial^2 E_1/\partial q_g^2) = E_C\,b_1(q_g,J)$. 
It is worth stressing that arbitrary small fluctuations determine 
$\delta(x) \neq 0$, therefore STIRAP may occur only via non-adiabatic 
patterns.   
In the same way also $\delta_p(x)$ depends on the derivatives
$(\partial^n E_2/\partial q_g^n)$.

In Fig.~\ref{fig:avetransf} we plot the populations histories 
$P_i(t)$ averaged over the fluctuations of $x$.  
These induce correlated fluctuations of both detunings and 
couplings $n_{ij}$. Device and bias
parameters correspond to a Quantronium biased slightly off-symmetry.
It is shown that low-energy fluctuations determine 
a $\sim 20 \%$ efficiency loss despite of the fact that protection 
from noise is greatly reduced. This is an interesting figure for 
superconducting nanodevices if we compare with the observed 
coherent population trapping of $\sim 60\%$ recently measured 
in phase-type devices~\onlinecite{ka:210-kelly-pappas-prl-cpt}. 
Moreover the population of the intermediate 
level is very small during the whole procedure, 
fulfilling the requirements for coherent population transfer. 

Such numerical evaluations are performed by using a 4-th order integration 
Runge-Kutta method for the solution of the 
ordinary differential equations. Convergence was tested 
down to a relative error lower
than $10^{-3}$, by adjusting both the integration step and the number of event
series. Integration over fluctuations was performed by  
a Montecarlo approach with up to 5000 samples in order to attain 
a relative error smaller than  $10^{-3}$.

Notice that the detunings depend on a single random variable $x$, 
therefore their fluctuations are correlated. In particular,  
charge noise determines anticorrelated 
fluctuations of effective detunings in CPB's, as it is clear from the 
spectrum (Fig.~\ref{fig:spectrum}.a).
This implies that non-ideal STIRAP may occur only via the typical 
LZ patterns~\cite{ka:212-falci-phscripta-opendriven} 
shown in Fig.~\ref{fig:stirap-zener}.a. 

\begin{figure}[t!]
\centering
\resizebox{0.4\textwidth}{!}{
\includegraphics{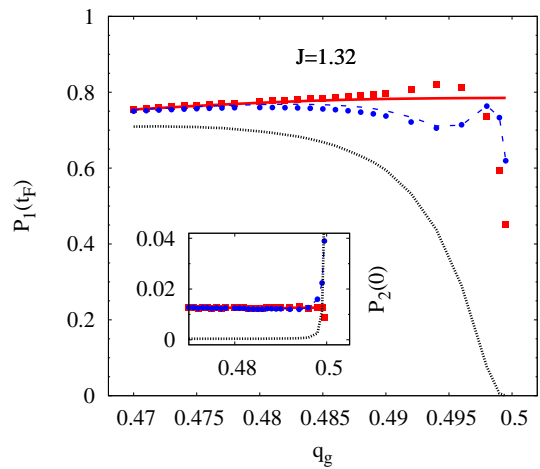}}
\caption[Effect of slow noise in the Quantronium]{
(color online) Efficiency of STIRAP $P_1(t_f)$ as a function of the bias $q_g$ 
in the presence of low-frequency and BBCN for the 
Quantronium ($E_J/E_C=1.32$). Here $\Omega_0 T = 15$  
$\sigma_x = 0.004$, $\nu_{01} = 600 MHz$. 
Upper curves show effects of low-frequency noise, whereas the 
lower curve (black dashed) includes also high-frequency noise. 
Low-frequency noise is analyzed by adding different components, namely 
linear and quadratic correlated fluctuations of detunings 
(red solid curve, and red squares), linear and quadratic fluctuations of 
$n_{02}$ (blue solid curve, and blue dots). For off-symmetry bias ($q_g<0.9$), 
only linear detuning noise is important (see App.~\ref{sec:more-fluct-eff} for
the behavior near  $q_g<1/2$). 
In the inset  the population $P_2(0)$ at intermediate times is shown.
\label{fig:110807-a}
}
\end{figure}

Notice that in the regime of Fig.~\ref{fig:avetransf} 
fluctuations of the couplings $n_{ij}$ could have been neglected. 
Indeed, they can be estimated from Fig.~\ref{fig:spectrum}.b. For 
instance, for $J \sim 1$ and $q_g < 0.49$ fluctuations of the amplitude 
of the pump pulse are $\sigma_{p} \sim a_{02} \, \sigma_x \Omega_0$, 
where $a_{02}= \partial n_{02}/\partial q_g$, therefore 
$\sigma_{p} \ll \Omega_0$. Numerical results 
(Figs.~\ref{fig:110807-a},\ref{fig:efficiency-other}) actually 
confirm that fluctuations $n_{ij}$ yield at most corrections, and  
moreover when they are appreciable STIRAP does not work due to 
the combined effect of high-frequency noise 
(see \S\ref{sec:stirap-quantronium}).

The above observation implies that for practical purposes 
efficiency can be discussed entirely in terms of the sensitivity
to detunings~\cite{ka:211-lacognata-ijqi-stirapcpb}.  
Diagrams in the $(\delta,\delta_p)$ plane 
(Fig.~\ref{fig:stirap-diamond}) can be used to understand the effect of 
low-frequency noise. Correlated stray detunings 
in the CPB are there represented by lines in the $\delta-\delta_p$ plane, 
which are straight lines for linear noise. We draw the segment corresponding 
to fluctuations $x \in [-\sigma_x,\sigma_x]$ (here $\sigma_x=0.004$),   
for each bias point ($q_g= 0.47,0.49$ are shown, the slope increasing 
by approaching $q_g=1/2$).   
If segments lie inside the light zone the efficiency is large. 
It is seen that efficient STIRAP requires small fluctuations 
$|\delta| < 0.1\,\Omega_0$ but the 
large anticorrelated $|\delta_p| \le 2.5 \,\Omega_0$ is tolerable.

\subsection{Effect of high-frequency noise}
High-frequency noise is studied by solving 
the quantum-optical ME in the rotating 
frame~\cite{ka:199-kuhn-singlephoton-APB}
$\dot{\rho} = \frac{i}{\hbar}[\rho,\tilde{H}] - \mathrm{D} \rho $,
where $\rho$ is the density matrix and $\tilde{H}$ is the 
Hamiltonian~(\ref{eq:lambdaconfig}). 
The structure of the dissipator $\mathrm{D} \rho$ 
in the basis of the diabatic states $\{|\phi_i\rangle\}$ 
reads~\cite{ka:209-siebrafalci-prb}
\begin{equation}
\label{eq:lindblad-decoh}
(\mathrm{D} \rho)_{ij} = \frac{\gamma_i+\gamma_j}{2}
                                          \rho_{ij}
         -\delta_{ij}\sum_{k\neq i} \rho_{kk}
                                     \gamma_{ik}
         + (1-\delta_{ij})\tilde{\gamma}_{ij} \rho_{ij}
\end{equation}
The first two terms describe emission and absorption of energy and the
associated secular dephasing: $\gamma_{ij} = \gamma_{j \to i}$ 
are transition rates between diabatic states, 
and  $\gamma_i=\sum_{k\neq i} \gamma_{ki}$ are the total decay rates of states
$\ket{\phi_i}$. At low temperature in an undriven system 
only rates of spontaneous emission between diabatic 
states are non negligible. In AC driven systems 
rates describing environment-assisted absorption are also nonzero, when  
the corresponding field is switched on~\cite{ka:195-geva-kosloff-jchemph-GME}. 
Finally the dissipator may include pure dephasing rates 
$\tilde{\gamma}_{ij}= \tilde{\gamma}_{ji}$.
 
\begin{figure}[t]
\centering
\resizebox{0.45\textwidth}{!}{\includegraphics{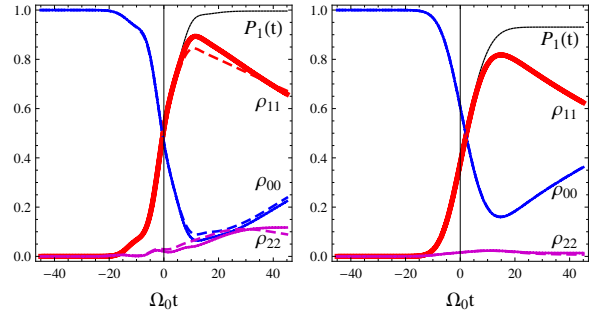}}
\caption{(color online) Population histories $\rho_{ii}(t)$ 
in the presence of high-frequency noise ($P_1$ is in absence), at resonance 
($\delta = \delta_p = 0$, left panel) 
and for finite anticorrelated detunings 
($\delta =0.05$, $\delta_p = -25 \delta$, 
right panel), for $\kappa=1$.
Solid lines are obtained by inserting only $\gamma_{01}=1/T_1$ in 
Eq.(\ref{eq:lindblad-decoh}), describing  relaxation 
$1 \to 0$ only, and the associated secular dephasing. We have chosen a rather 
large $\gamma_{01}/\Omega_0=0.01$ to emphasize the effect. 
Dashed lines take into account all the other low-temperature emission and 
drive-induced absorption channels (the chosen rates overestimate these
processes), which are seen to have a limited impact on the efficiency. 
Physical scales for 
$\Omega_0=3.46 \times 10^8 \mathrm{rad/s}$ (the value we use
for the Quantronium at $q_g = 0.48$) are
$T \approx 43\,\mathrm{ns}$, for the overall protocol $T_T \approx 290\,
\mathrm{ns}$ and the chosen $T_1 \approx T_T$. 
\label{fig:stirap-quantum-noise-0}}
\end{figure}
In quantum optical systems STIRAP connects two ground states, 
$\gamma_{01}=\gamma_{10}=0$. 
Therefore as long as population in 
$\ket{\phi_2}$ is small all the transition rates 
act on depopulated states, and it is known that 
they practically do not affect population transfer. 
Instead in superconducting nanocircuits 
the decay channel $\gamma_{01}$ is active. Therefore, 
we expect that $\gamma_{01}$ is the main source of efficiency 
loss due to processes involving energy exchange with the environment. 
This is indeed the qualitative conclusion suggested by the results in  
Fig.~\ref{fig:stirap-quantum-noise-0}.  

To clarify the physical picture 
in Fig.~\ref{fig:stirap-quantum-noise-0}
we study separately the impact of adding decay channels.  
First we consider only spontaneous decay in the first doublet.
We take $\gamma_{01}/\Omega_0=0.01$, which is 
a rather large value used to emphasize the effects  
and we study population histories (solid lines $\rho_{ii}$). 
We find $\rho_{11}(t) \approx P_1(t)\,\mathrm{e}^{- \gamma_{01}(t-t_i)} $, 
where $P_1(t)$ is the population in absence of noise, therefore 
this channel mainly determines the simple population loss $1 \to 0$ 
when the target state is populated. 
It also determines a nonvanishing population
$\rho_{22} \neq 0$ which indicates 
detrapping from the dark state due to loss of coherence.

Adding all the other decay channels (dashed lines) 
produces minor modifications of this picture 
(Fig.~\ref{fig:stirap-quantum-noise-0} left panel) for fields at resonance.
No modification at all occurs  
for nonvanishing detunings, mimicking 
low-frequency fluctuations. The reason is that in this latter case 
$\gamma_{01}$ does not determine 
substantial detrapping, and population of $\ket{\phi_2}$.

In detail results of Fig.~\ref{fig:stirap-quantum-noise-0} were obtained
by using rates for the other decay channels which overestimate 
unwanted processes, namely
$\gamma_{12}= 2\,\gamma_{01}$, whereas $\gamma_{02}= 0.2 \,\gamma_{01}$
(accounting for the suppression by selection rules). 
Notice that we do not take into account the fact that 
these emission rates become smaller when the drive amplitudes 
$\Omega_k(t)$ are large enough, as resulting from 
the generalized (Bloch-Redfield) ME
for AC driven systems undergoing Rabi 
oscillations~\cite{ka:195-geva-kosloff-jchemph-GME}. This latter 
approach shows that also  
in the weak damping Rabi regime ($T\Omega_k(t) \gg 1$ and 
$\delta_k \ll \Omega_k(t)$) field-induced absorption sets in, even at 
low temperatures. 
We take into account this channel 
phenomenologically, letting 
$\gamma_{21}(t) =
\gamma_{12}/4\,[1-\delta_s/(\sqrt{\delta_s^2+\Omega_s^2(t)})]^2 
\,g[\Omega_s(t)T]$, where 
$g(x) \approx 1$ only for $x \gg 1$ accounts for the requirement that 
field induced processes set in for underdamped Rabi oscillations. We 
used a similar expression for $\gamma_{20}(t)$.

Notice that while secular dephasing is taken into account in 
Fig.~\ref{fig:stirap-quantum-noise-0} we did not include 
Markovian pure dephasing rates, $\tilde{\gamma}_{ij}=0$. Indeed, we 
argue that pure dephasing comes mainly from low-frequency (non-Markovian) 
noise accounted for by classical fluctuations of $x$. In the next section
we study the combined effect of high and low-frequency noise. 
We will discuss different models of pure dephasing 
in Sec.~\ref{sec:comparison}. 
We finally mention that, for charge-phase CPB 
it has been shown that operating at $q_g = 0.48$ already provides 
sufficient coupling $n_{02}$ to observe STIRAP in the presence of the 
Markovian component of noise~\cite{ka:208-mangano-epj}. 
 
\subsection{Combined effect of low and high-frequency noise}
\label{sec:stirap-quantronium}
The main conclusion of the last two sections is that 
the leading effects reducing coherent population transfer 
in nanodevices essentially involve decoherence of the first doublet. 
Another detrimental effect is that 
coupling to the pump pulse may 
be too weak due to (approximate) parity selection rules at (near)
the symmetry point.  

With this in mind we investigate 
the interplay of low and high-frequency 
fluctuations, for $q_g \le 1/2$ in a charge-phase CPB. Indeed we 
will argue in the next section that STIRAP can be observed only in this 
regime. Here we consider a case-study device as the 
Quantronium, where noise in the first doublet has been well characterized. 
We take the value $T_1 = 1\,\mu\mathrm{s}$ which is achievable in the class
of CPB devices at $q_g=1/2$, and neglect its weak dependence on the 
bias~\cite{ka:205-ithier-prb}. 

Results are summarized in Fig.~\ref{fig:110807-a} where the efficiency is 
plotted against the bias $q_g$, showing the impact of adding various 
low-frequency and high-frequency decoherence channels. Curves refer to the 
same $\Omega_0T=15$, which guarantees adiabaticity for ideal STIRAP.
It is apparent the different behavior sufficiently far ($q_g < 0.49$) 
and close ($q_g \approx 0.5$) to the symmetry point. 

For off-symmetry bias it is possible to 
observe STIRAP despite of the reduced protection from low-frequency noise. 
In this regime low-frequency noise is the main source of efficiency loss
allowing a population transfer close to $\sim 80 \%$. Notice that only linear 
fluctuations of the detunings are important: indeed Fig.~\ref{fig:110807-a} 
shows that accounting for the whole structure of 
low-energy fluctuations yields basically the same result, as 
the dependence of $E_{10}$ and $n_{02}$ on $q_g < 0.49$ would a priori
suggest. Efficiency is reduced to $\sim 70 \%$
when also effects of high-frequency noise are taken into account
(black dashed curve Fig.~\ref{fig:110807-a}).

Instead by approaching $q_g=1/2$, while low-frequency fluctuations 
would still allow for some population transfer, the 
interplay with high-frequency noise, mainly 
due spontaneous decay 
$\ket{\phi}_1 \to \ket{\phi_0}$,  leads  
to the suppression of the efficiency 
(solid curve in Fig.~\ref{fig:110807-a}).
Actually, in this regime the description of the effect of low-frequency 
fluctuations is more complicated spoiling the simple picture based on 
sensitivity to detunings. We discuss in App.~\ref{sec:more-fluct-eff} 
the whole information contained in Fig.~\ref{fig:110807-a}. The 
main point is the observation that even if the device is biased at $q_g=1/2$, 
fluctuations still allow for a nonvanishing pump coupling 
despite of the parity selection rule.
The reason why high-frequency noise suppresses the efficiency is 
understood by recalling that STIRAP requires large pulse area, 
$\Omega_p(q_g+x) T \gg 10$. Since close enough to the symmetry point 
$\Omega_p$ becomes small, larger and larger $T$ are needed 
which eventually exceed by far $T_1$.
This mechanism explains the fact that the loss of 
efficiency due to high-frequency noise  
(Figs.~\ref{fig:110807-a},\ref{fig:efficiency-other}) 
appears to depend strongly on $q_g$,  
even if we neglected the (in any case weak) dependence of the rates
$\gamma_{ij}$ on the bias.

The population $P_2(0)$ of the intermediate level 
during the adiabatic passage phase remains small in the 
presence of BBCN 
(inset of Fig.~\ref{fig:110807-a}). This is an essential requirement for 
success and applications of the protocol, and completes the  statement that 
STIRAP should be observable in charge-phase CPB's. 

\begin{figure}[t!]
\centering
\resizebox{0.45\textwidth}{!}{\includegraphics{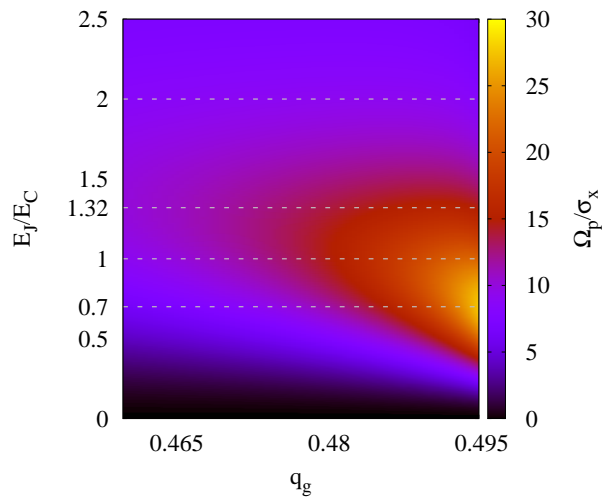}} 
\caption{(color online)
The figure of merit $\Omega_p/\sigma_\delta$ is plotted in the 
$(q_g,E_J/E_C)$ plane. We have chosen $\sigma_x=0.004$
and $\Omega_0$ produced by an external field, which would determine 
Rabi oscillations with  $\nu_R = 600 \,\mathrm{MHz}$ in the first doublet.
The analysis is valid far enough from the charge-parity symmetry point, 
which is not an interesting regime since the efficiency is suppressed. 
Dashed lines correspond to the values of $E_J/E_C$ checked in this paper
(Figs.~\ref{fig:110807-a} and \ref{fig:efficiency-other}).
\label{fig:linewidth}
}
\end{figure}

Finally we mention that working with larger asymmetry, while providing 
a stronger pump coupling (see Fig.~\ref{fig:spectrum}) enhances the effect 
of low-frequency noise, reducing the overall efficiency. This is 
apparent from the trend in Fig.~\ref{fig:110807-a} (see also 
Fig.~\ref{fig:linewidth}), indicating that optimization of strong enough 
pump coupling and protection from low-frequency noise is a key issue 
for the implementation of a $\Lambda$ system.

\section{Optimal design of the device}
\label{sec:optimal-design}
Efficiency of population transfer may be 
improved by optimizing the parameters of the protocol. In the last section we have shown that, due to the combined effect of the approximate symmetry and of spontaneous decay, 
efficiency is large enough only if the device is biased slightly 
away from the symmetry point. In this  
section, we argue that in this regime one should mainly optimize 
the tradeoff between coupling of the pump pulse and 
energy fluctuations of the lowest doublet of the device, due to 
low-frequency noise.  
Indeed the relevant figure of merit turns out to be
\begin{equation}
{2 E_C \langle n_{02} \rangle \over \sigma_\delta} 
\,\propto \, 
{ \Omega_p^{max}  \over \sigma_\delta} 
\label{eq:optimization}
\end{equation}
where $\sigma_\delta = 
\sqrt{A_1^2 \sigma_x^2 + {1 \over 2} B_1^2 \sigma_x^4}$
are the fluctuations of $\delta(x)$.
Efficient population transfer requires 
large enough values of this quantity,
which depends on $q_g$ and on $J$, via $\langle n_{02} \rangle$
and $(A_1,B_1)$.  This allows to choose convenient
design and operating conditions
(see Fig.\ref{fig:linewidth}). 

\begin{figure}[t!]
\centering
\resizebox{0.33\textwidth}{!}{\includegraphics{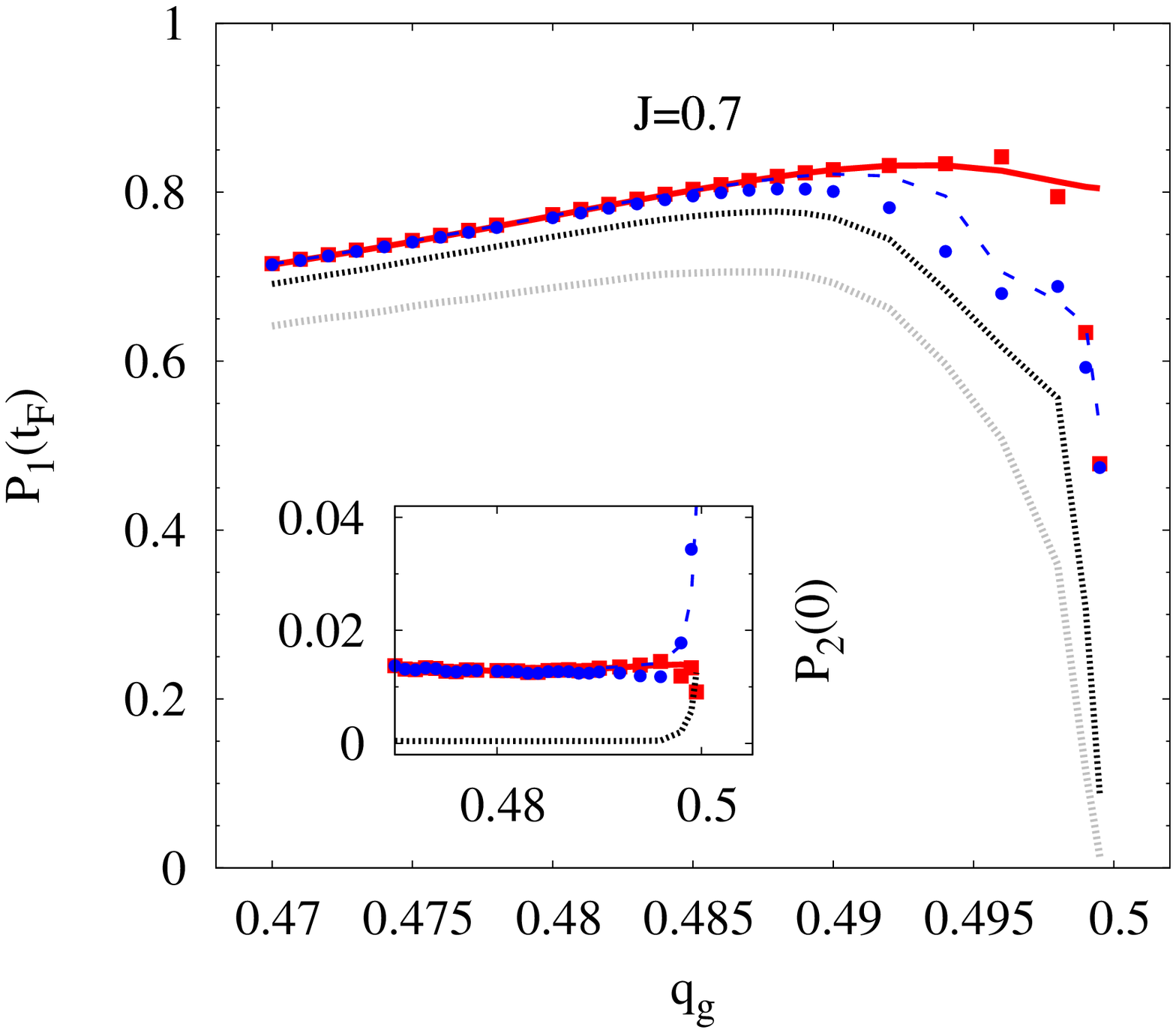}}
\resizebox{0.33\textwidth}{!}{\includegraphics{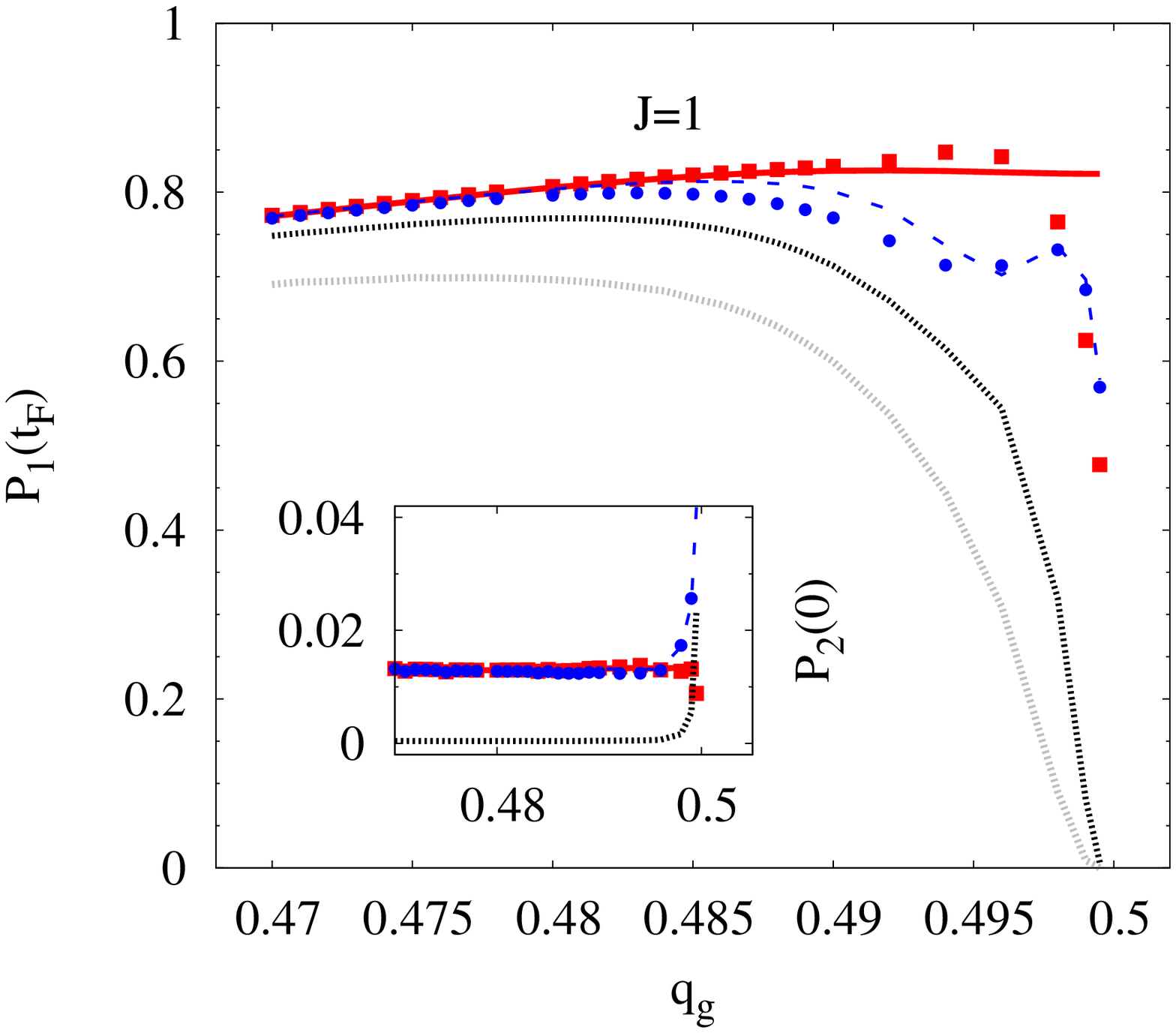}}[
\resizebox{0.33\textwidth}{!}{\includegraphics{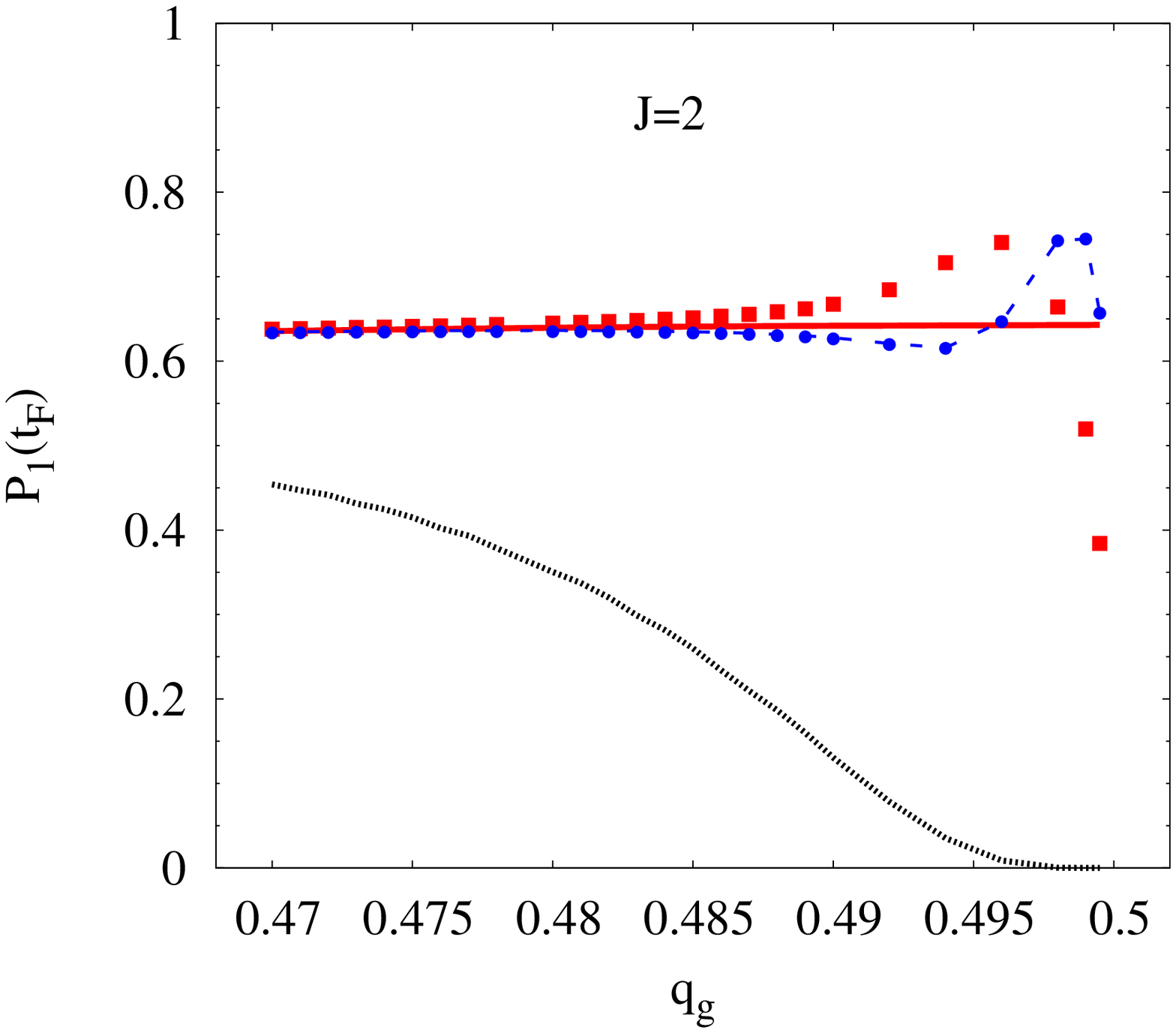}}
\caption{(color online) Efficiency $P_1(t_f)$ vs. bias $q_g$ 
for $J=0.7,1,2$ (see Fig.~\ref{fig:linewidth}). Parameters are the same 
as in Fig.~\ref{fig:110807-a}, where $T_1=1000\,\mathrm{ns}$ 
(black short-dashed curves). In the two upper panels efficiency for 
smaller $T_1=500\,\mathrm{ns}$ (gray short-dashed curve) is also shown.
\label{fig:efficiency-other}
}
\end{figure}

The criterion is clear from heuristic grounds, but can also be 
justified starting from an estimate of the linewidth for population transfer
at finite $\delta$. To this end we generalize an argument given by Vitanov 
et al.~\cite{kr:201-vitanov-advmolopt}. They 
noticed that even if states of the 
adiabatic basis $\{\ket{D},\ket{\pm}\}$ are not anymore 
instantaneous eigenstates, 
still $\ket{D}$ provides a connection between the diabatic states 
$\ket{\phi_0}$ and $\ket{\phi_1}$. Then it is argued that 
efficiency loss depends on processes triggering transitions from $\ket{D}$ 
to $\ket{\pm}$. These are due to non vanishing off diagonal entries of the 
Hamiltonian in the adiabatic basis, which are proportional to 
$\delta$. Therefore, if $\delta \gg \min |\epsilon_{\pm}|$ population transfer 
does not occur. This condition implies that 
for $\delta_p = 0$ the linewidth scales 
linearly with the amplitude of the fields~\cite{kr:201-vitanov-advmolopt},  
$\delta_{1 \over 2} = d(\tau) 
\sqrt{(\Omega_p^{max})^2+(\Omega_s^{max})^2}$.  
In our case stray detunings $\delta(x)$ 
and $\delta_p(x) \neq 0$ are anticorrelated, therefore leakage from $\ket{D}$ 
occurs during the pump phases (see  Fig.~\ref{fig:stirap-zener}.a). 
Moreover, from Fig.~\ref{fig:stirap-diamond} we see 
that it is substantial only when $\delta_p > 2 \Omega_0$. 
In this regime the relevant condition 
$\delta_{1 \over 2}= |\epsilon_-|$ is an equation whose solution 
can be still written as  $\delta_{1 \over 2} \approx d^\prime(\tau,\kappa) 
\,\Omega_p^{max}$. 
Asking that fluctuations of $\delta$ do not destroy the efficiency 
means that we need $\sigma_\delta \ll \delta_{1 \over 2}$. 
Therefore, we need large values of the parameter
$\delta_{1 \over 2}/\sigma_\delta \propto \Omega_p^{max}/\sigma_\delta$, 
which justifies the figure of merit defined in 
Eq.(\ref{eq:optimization}). Our derivation does not take into account 
fluctuations of the matrix elements, since they are negligible in the regime 
where STIRAP could work. For the same reason we did not include in 
Fig.~\ref{fig:linewidth} the region near
$q_g = 1/2$, since in this regime STIRAP is in any case prevented 
by spontaneous decay,
due to the too low achievable values of $\Omega_0 T_1$.

We check the optimization suggested by Fig.\ref{fig:linewidth} by looking 
at STIRAP for different values of $J$. It is seen that
proper fabrication parameters allows to obtain larger efficiency 
($J=0.7$ in Fig.\ref{fig:efficiency-other}a and 
$J=1$ in Fig.\ref{fig:efficiency-other}b). Instead
for larger values of $J$, as in the 
Transmon~\cite{ka:207-koch-pra-transmon} design, pump coupling 
is insufficient even if protection against noise is much better. 
In the opposite limit of charge qubits
$E_J/E_C \ll 1$, the efficiency is also small because of both small coupling 
and reduced protection from noise. This latter strongly suppresses 
population transfer also for a bias $q_g$ far-off symmetry, despite the 
coupling to the field increases. Notice that the dependence of the 
efficiency on the parameter $J$, besides providing 
prescriptions for the fabrication, can also be checked by on-chip tuning of 
$E_J$ via an external magnetic flux $\Phi_g$ (see Fig.~\ref {fig:quantronium}).

\section{Comparison of different mechanisms of dephasing}
\label{sec:comparison}
Studying low-frequency noise in nanodevices by a non-Markovian
model is necessary to explain quantitatively 
striking experimental features observed in quantum bits, 
as the peculiar non-exponential initial 
decoherence~\cite{ka:205-ithier-prb,ka:211-bylander-natphys,ka:212-chiarello-njp}. 
Moreover, this approach provides valuable additional information 
as relations between effects of noise for different bias point~\cite{ka:205-ithier-prb,ka:211-bylander-natphys,ka:212-chiarello-njp} and different device 
design~\cite{ka:212-sank-martinis-prl-fluxnoise}, which are
uniquely explained by the parametric dependence of the energy spectrum.
We stress that such a picture is entirely due to non-Markovianity of BBCN. 

\begin{figure}[t]
\centering
\resizebox{0.45\textwidth}{!}{\includegraphics{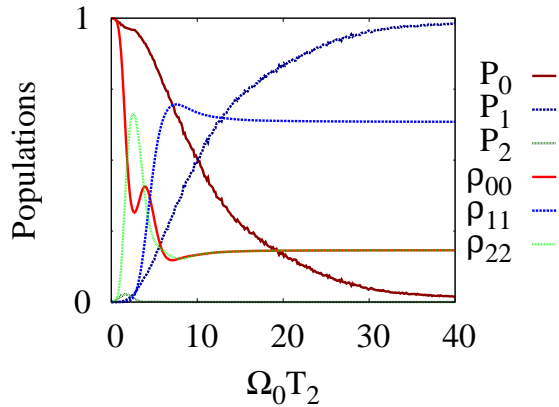}} 
\caption{(color online) Efficiency of STIRAP (final populations) 
as a function of the drive amplitudes $\Omega_0$. We compare the case 
of Markovian ($\rho_{ii}$) pure 
dephasing~\cite{ka:204-ivanov-pra-stirapdephasing} with the 
non-Markovian ($P_i$) model studied here. 
In both cases we let $T=T_2=57\, \mathrm{ns}$, which for non-Markovian noise 
is obtained by taking $\sigma_x = 0.004$ in a device with $J=1.32$ at 
$q_g=0.48$. It is seen that the effects of non-Markovian dephasing can be 
attenuated and suppressed by using larger $\Omega_0$, whereas for Markovian
noise STIRAP, when effective, does not depend on $\Omega_0$.
\label{fig:motional-narrow}}
\end{figure}

Therefore, this work complements previous studies in the quantum optics 
realm where typically the Markovian ME is used.
In this latter approach pure dephasing is studied by considering only  
nonvanishing dephasing rates $\tilde{\gamma}_{ij}$ 
in the dissipator Eq.(\ref{eq:lindblad-decoh}), 
instead of the static fluctuations considered in this paper. 
Pure dephasing in the Markovian ME was studied by
Ivanov et al.~\cite{ka:204-ivanov-pra-stirapdephasing}, 
who derived an adiabatic solution of the Liouville equation  
interpolating between the coherent and the incoherent limit. 
They predicted striking behaviors as a function of the control parameters, 
deriving several analytic results, which have been numerically checked. 
In particular for Gaussian pulses, populations at the end of the protocol 
were found to be
\begin{equation}
\label{eq:ivanov}
\begin{aligned}
\rho_{11}(\infty) &= {1 \over 3} + {2 \over 3} \,
\mathrm{e}^{-3 \tilde{\gamma}_{01} T^2/(8 \tau)}
\\
\rho_{00}(\infty) &= \rho_{22}(\infty) = 
{1 \over 3} - {1 \over 3} \,
\mathrm{e}^{-3 \tilde{\gamma}_{01} T^2/(8 \tau)}
\end{aligned}
\end{equation}
Notice that in this approximation 
the efficiency is determined by the dephasing rate of the 
lowest doublet only, $\tilde{\gamma}_{01}$.  The conclusion that 
other dephasing channels are less relevant (actually for 
$\tilde{\gamma}_{12},\tilde{\gamma}_{02} \gg \tilde{\gamma}_{01}$ 
some dependence appears 
in the numerical solutions of the ME) agrees qualitatively with 
our results with the static fluctuator model.
On the contrary, the other striking feature of Eq.(\ref{eq:ivanov}), 
namely that losses due to dephasing are independent on the peak Rabi 
frequencies, does not hold for low-frequency noise.
Following Ref.~\onlinecite{ka:204-ivanov-pra-stirapdephasing} 
we plot in Fig.~\ref{fig:motional-narrow}  populations 
$\rho_{ii}(\infty)$ obtained numerically from the Markovian ME, using  
fixed $T=T_2^*=1/\tilde{\gamma}_{01}$,   
for increasing pulse amplitude $\Omega_0$. 
We compare them with the populations
$P_i(\infty)$ for the BBCN non Markovian model, where 
linear fluctuations of the detunings are considered 
such that $\sigma_x= \sqrt{2}/(A_1 T_2^*)$, which yield the 
same $T_2^*$ in the qubit dynamics. 
It is seen that efficiency for BBCN depends on 
$\Omega_0$ and improves for increasing values. 

Dependence on $\Omega_0$ is a natural consequence of non-ideal 
STIRAP occurring via LZ patterns determined by 
low-frequency noise. Markovian noise cannot account for 
this scenario. The situation here is reminiscent of dynamical 
decoupling~\cite{ka:viola-bb} 
which eliminates dephasing for $1/f$ noise sources, 
as the effect of a strong continuous AC fields also 
does~\cite{ka:205-facchipascazio-pra-threestrategies}. 

Another difference between Markovian and non-Markovian dephasing is that 
this latter practically does not
populate  the intermediate level $\ket{\phi_2}$, 
although it decreases the transfer efficiency. This is another indication
of the reduced sensitivity of the protocol to low-frequency noise. 
On the contrary, sensitivity of $\rho_{22}$ to Markovian noise 
is substantial and could give direct informations on 
$\tilde{\gamma}_{01}$, as seen from Eq.(\ref{eq:ivanov}). 
This observation is reminiscent of the  
proposal of Ref~\onlinecite{ka:204-murali-orlando-prl-eit} of
using EIT to probe decoherence of a phase qubit
based on a SQUID nanodevice. Having in mind realistic noise spectra
it is likely that the contribution of intermediate frequencies may 
determine effects similar to Markovian dephasing. Therefore, 
cross checking measurement of decoherence of two and three-level  
dynamics could give valuable spectral-resolved information on 
the environment.

We stress the striking implication of non-Markovianity of the noise,  
namely correlations between fluctuations of the detunings, 
entirely determined by the parametric dependence of the energy spectrum. 
Effects of time-correlated (Ornstein Uhlembeck) phase noise in optical 
systems were studied by Monte Carlo simulations
in Ref.~\onlinecite{ka:202-yatsenko-pra-stirapphasenoisecorr}, where 
the regime of partially correlated  $\delta_p$ and $\delta_s$ 
was addressed. In nanodevices the situation is different 
since we have strongly anticorrelated (or correlated) 
stray detunings $\delta$ and $\delta_p$.
It would be interesting to investigate dynamic phase diffusion 
also in this case.
\section{Conclusions}
\label{sec:conclusion}

In this paper we have studied the combined effect of low-frequency and 
high-frequency charge noise on the coherence of a CPB operated as a 
three-level artificial atom in Lambda configuration. 
Observation of STIRAP
should be possible in devices within present fabrication standards, 
provided both design and operating conditions are carefully chosen. 

We have shown that efficient population transfer requires 
optimizations of the tradeoff between
large enough pump coupling and the implied larger 
sensitivity to low-frequency noise. To this end the CPB should be 
biased slightly off-symmetry in a region where 
low-frequency fluctuations of the energy spectrum are linear in the 
fluctuations $x$ of the control parameter $q_g$. 

We have shown that the noise is conveniently 
analyzed by mapping it onto fictitious correlated fluctuations of the 
detunings (see Fig.~\ref{fig:stirap-diamond}). This simple picture 
emerges because, despite of the complications brought by the  
multidimensional space of parameters,
the efficiency for STIRAP is shown to depend essentially on 
noise channels relative to the trapped subspace only. 
The relevant channels can be fully characterized by operating the 
nanodevice as a qubit, as in Refs.~\onlinecite{ka:202-nakamura-chargenoise,ka:205-ithier-prb,ka:211-bylander-natphys,ka:212-chiarello-njp,ka:212-sank-martinis-prl-fluxnoise}.

We have found that the tradeoff is 
summarized by a single figure of merit, given in Eq.(\ref{eq:optimization}),
which indicates favorable conditions for observation of STIRAP.
Its remarkable dependence on features of the three-level spectrum 
of the device (energy correlations, symmetries) 
suggests that band structure engineering may play a 
key role in determining optimal design solutions. This analysis, 
together with other already available tools, as improvements in materials 
and control circuits, besides a systematic investigation of 
parameters and pulses crafting, 
guarantees room for further improvement of the efficiency.

In this work we did not consider other noise sources (as the readout circuit
or critical current noise), which are possibly coupled to the device 
in channels ``orthogonal'' to the drive. This is because in the 
successful regime for STIRAP they lead to minor effects 
in CPB's~\cite{ka:205-ithier-prb}. They can be easily accounted 
for by a slight generalization of our approach, 
allowing for independent noise sources. Notice that each noise source 
could determine its own correlations of $\delta$ and $\delta_p$.

We remark that the physical 
picture emerging from this work applies to the whole class 
of superconducting nanocircuits, used so far 
for implementing quantum bits~\cite{kr:208-clarke-wilhelm-nature-squbit}. 
Our full analysis applies to flux-qubits~\cite{flux-etc} 
where a coordinate-parity selection 
rule holds~\cite{ka:205-nori-prl-adiabaticpassage} 
and a symmetry point exists, except that  
two orthogonal noise sources (flux and critical current 
plus charge~\cite{ka:211-bylander-natphys})
should be taken into account for accurate predictions.
It also applies to phase-qubits~\cite{phase-etc} where only 
linear fluctuations are important~\cite{ka:203-martinis-prb-noise}, 
but detunings are differently correlated. 
In all these devices the figure of merit analogous 
to that of Eq.(\ref{eq:optimization}) can be 
used to characterize 
the effect of low-frequency noise versus efficient coupling.

A natural extension of our work is the investigation of 
dynamic diffusion for correlated phases/detunings in the experimentally 
relevant case of $1/f^\alpha$ noise, and to which extent the dependence 
on the drive intensity of the resilience to low-frequency noise can be
used for some effective dynamical decoupling.
Moreover it has been pointed out that 
the presence of one or few more strongly coupled fluctuators may  
deteriorate the efficiency of ideal 
STIRAP~\cite{ka:212-vogt-prb-stirap} and it would be interesting
to extend the investigation to the LZ scenario. Finally 
circuit-QED~\cite{ka:204-wallraff-nature-cqed} based architectures 
are natural candidates for the implementation of STIRAP with quantum fields, 
the physics related to BBCN~\cite{ka:212-paladino-physscr-purcell} 
must be studied in this broader scenario.

\appendix
\section{More on the CPB}
\subsection{Driven three-level effective Hamiltonian}
\label{app:cpb-effhamil}
Manipulation of the quantum state is performed by adding to the DC 
part of the gate voltage 
AC microwave pulses with small amplitude, 
$q_g \to q_{g}+q_c(t)$. The resulting 
Hamiltonian can be written as 
\begin{equation}
\label{eq:quantronium_hamilt-ac-2}
H(t) = H_0(q_g) + A(t) \, \hat{n} 
\end{equation}
where  $A(t)=-2E_C q_c(t)$. The effective three-level artificial atom Hamiltonian is 
obtained by projecting $H(t)$ onto the subspace spanned by the three lowest energy 
eigenvectors $|\phi_i\rangle$, $i=0,1,2$ of $H_0(q_g)$ 
\begin{equation}
\label{eq:quantronium_hamilt-ac-trunc}
H(t) = \sum_i E_i 
|\phi_i\rangle \langle \phi_i|
+ A(t) \, \sum_{ij} n_{ij} \,|\phi_i\rangle \langle \phi_j|
\end{equation}
where $n_{ij}=\langle \phi_i | \hat{n} |\phi_j \rangle$.
The STIRAP protocol can be carried out
if we let $A(t)={\cal A}_s(t)\cos{\omega_s t}+{\cal A}_p(t)\cos{\omega_p t}$. 
We then perform the RWA, by retaining only quasi resonant 
off-diagonal and corotating terms of the drive, 
which simplifies to
\begin{equation}
\label{eq:adiab-pass-charging-rwa}
\begin{aligned}
A(t) \hat{n} \,\to\, 
H_{RW} (t)  &
= {1 \over 2}\,\big[
n_{12} \, {\cal A}_s(t) \,
\mathrm{e}^{i \omega_{s} t}  |\phi_1\rangle \langle \phi_2|
\\
+& 
n_{02} \, {\cal A}_p(t) \,
\mathrm{e}^{i \omega_{p} t}  |\phi_0\rangle \langle \phi_2|
\big]
+ \mbox{h.c.}  
\end{aligned}
\end{equation}
Finally, the Hamiltonian 
is transformed to the doubly rotated frame, 
at angular frequencies $\omega_s$ and $\omega_p$ via the transformation
$U_{rf} = \exp[i (\omega_s \ket{\phi_1}\bra{\phi_1} + 
\omega_p \ket{\phi_0}\bra{\phi_0}) t]$.  
This yields an effective Hamiltonian $\tilde{H}(q_g)$ with the structure of 
Eq.(\ref{eq:lambdaconfig}), 
implementing the $\Lambda$ configuration.
Notice that $n_{ij}=\langle \phi_i | \hat{n} |\phi_j \rangle$  play the same
role of the dipole matrix elements in the definition
Eq.(\ref{eq:rabi-frequencies}) of  the Rabi frequencies.

\subsection{Charge-parity symmetry and selection rules}
\label{app:cpb-parity}
Charge parity is a possible symmetry of wavefunctions in charge space
which emerges because of the discrete nature of the momentum. 
Formally we introduce operators $\Pi_q=\sum_n \ket{q-n}\bra{n}$, 
which implement a reflection and then a translation in the charge space.
If the parameter $q$ is integer $\Pi_q$ always operates onto the same 
Hilbert state of discrete charges. It is easy to see that 
$$
\begin{aligned}
&\Pi_q^{-1} \big[\sum_n (n-q_g)^2 \ket{n}\bra{n} \big]\, \Pi_q =
\sum_n (n-q+q_g)^2 \ket{n}\bra{n}
\\
&\Pi_q^{-1}\big[\sum_n \ket{n}\bra{n \pm 1}\big] \Pi_q = 
\sum_n \ket{n}\bra{n \mp 1}
\qquad .
\end{aligned}
$$
Therefore, one can seek for the invariance of the family of Hamiltonians 
(\ref{eq:quantronium_hamilt}). Symmetry points are found for $q_g = q/2$,
where $H_0(q_g)$ is invariant with respect to 
$\Pi_{2q_g}$. Since $\Pi_q^2=\mathbbm{1}$,  
for symmetric $H_0$ eigenvalues can be chosen with a well 
defined charge parity $\Pi_{2q_g}\ket{\phi_j(q_g)}=(-1)^j \,\ket{\phi_j(q_g)}$
and parity selection rules hold such that 
for states of different parity
charge matrix elements vanish, 
$\bra{\phi_j} n  \ket{\phi_i} =0$.

\section{Fluctuational behavior near the symmetry point}
\label{sec:more-fluct-eff}
We give a more detailed account on the effects of low-energy fluctuations 
close to $q_g=1/2$, displayed in 
Figs.~\ref{fig:110807-a},\ref{fig:efficiency-other},  
and on how they combine with high-frequency noise.
On approaching $q_g=1/2$, fluctuations of the detunings turn from 
linear to quadratic. These fluctuations alone 
(thick red squares in the figures)
would determine  a nonmonotonic behavior of the efficiency 
on approaching the symmetry point. 
Indeed for 
$0.49 \lesssim q_g \lesssim 0.495$ fluctuations for $x>0$ yield smaller 
stray detuning $\delta(q_g+x)$ than in the linear approximation, and the 
efficiency increases. However,   
approaching the symmetry point fluctuations 
$\langle |\delta| \rangle$ exceed the 
linewidth $\delta_{1 \over 2}$. Indeed, since this latter scales with 
$\Omega_0$ (see sec.~\ref{sec:optimal-design}) 
and thus vanishes for $q_g \to 1/2$, we find 
$\langle |\delta| \rangle/\Omega_0 \to \infty$ and
the efficiency should eventually vanish. However in this regime 
also the effect of fluctuations of the couplings play a role. We study these 
fluctuations in linear and quadratic approximation, 
$n_{ij}(q_g+x) \approx n_{02}(q_g) + A_{02} x + {1 \over 2} B_{02} x^2$. 
It turns out that only fluctuations of $n_{02}$ are possibly relevant, 
and only in the regime where fluctuations of $\delta$'s 
are quadratic. However in this regime 
they spoil the picture based on fluctuations of $\delta$'s only. 
Indeed, for $0.49 \lesssim q_g \lesssim 0.495$ smaller values of 
$\Omega_p(q_g+x)$ for 
$x>0$ compensate the positive effect of smaller $\delta(q_g+x)$. On the
contrary, on approaching the symmetry 
point slow fluctuations of $n_{02}$ provide a nonvanishing coupling
which is enough to yield a nonzero efficiency. 
Notice that this is true  even at the nominal bias 
$q_g = 1/2$, where the selection rule is exact, since  
also in this limit $\langle |\delta/ n_{02}| \rangle$ 
is finite. 

In Figs.~\ref{fig:110807-a},\ref{fig:efficiency-other}  
we show the effect of linear (blue solid curve) and quadratic 
(blue dots) fluctuations of $n_{02}$. The two approximations differ for 
small $J$, indicating that the series expansion is likely
not accurate enough. This is not a problem for our description of 
STIRAP, which in this regime is anyway suppressed by spontaneous decay. 
For larger values ($J=1.32,2$) the series expansion is seen to be accurate.

\acknowledgments
One of us (GF) acknowledges E. Arimondo for having pointed out the 
mechanism of efficiency improvement by larger pulse amplitudes 
in non-ideal STIRAP. This work was partially supported by EU through 
Grant No. PITN-GA-2009-234970, and by MIUR through Grant. No.
PON02\_00355\_3391233, ``Tecnologie per l'ENERGia e 
l'Efficienza energETICa - ENERGETIC''.

\end{document}